\documentclass[useAMS,usenatbib,usegraphicx]{mn2e}

\usepackage{aas_macros}
\usepackage{amsfonts,amssymb}
\usepackage{amsmath}
\usepackage{wasysym}
\usepackage{url}

\newcommand{\mathsc}[1]{\text{\textsc{#1}}} 
\DeclareRobustCommand{\ion}[2]{%
\relax\ifmmode
\ifx\testbx\f 
{\mathbf{#1\,\mathsc{#2}}}\else
{\mathrm{#1\,\mathsc{#2}}}\fi
\else\textup{#1\,{\mdseries\textsc{#2}}}%
\fi}

\title[Large scale bias of the QSO LOS proximity
effect]{Large scale environmental bias of the QSO line of sight proximity
effect}
\author[A. M. Partl, V. M\"uller, G. Yepes, S. Gottl\"ober]
{A. M. Partl$^{1}$\thanks{E-mail: apartl@aip.de (AMP)},
V. M\"uller$^{1}$, G. Yepes$^{2}$, and S. Gottl\"ober$^{1}$\\
$^{1}$Leibniz-Institut f\"ur Astrophysik Potsdam, An der Sternwarte 16,
Potsdam, 14482, Germany\\
$^{2}$Grupo de Astrof\'isica, Universidad Aut\'onoma de Madrid, Madrid E-28049, Spain
}
\begin{document}

\date{Accepted 1988 December 15. Received 1988 December 14; in original
form 1988 October 11}

\pagerange{\pageref{firstpage}--\pageref{lastpage}} \pubyear{2009}

\maketitle

\label{firstpage}

\begin{abstract}
We analyse the overionisation or proximity zone of the intergalactic
matter around high-redshift
quasars in a cosmological environment. In a box of $64 \; h^{-1}
\textrm{ Mpc}$ base length
we employ high-resolution dark matter only simulations with $1024^3$
particles. For estimating the
hydrogen temperature and density distribution we use the effective equation of
state by \citet{Hui:1997uq}.
Hydrogen is assumed to be in photoionisation equilibrium with a model background flux 
which is fit to recent observations of the redshift dependence of the mean optical
depth and the transmission flux statistics. At redshifts $z= 3, 4, \mbox{and } 4.8$, we select 
model quasar positions at the centre of the 20 most massive halos and 100 less massive
halos identified in the simulation box. From each assumed quasar position
we cast 100 random lines of sight for two box
length including the changes in the ionisation fractions by the QSO flux field
and derive mock Ly$\alpha$ spectra. The proximity effect describes
the dependence of the mean
normalised optical depth $\xi=\tau_{{\rm eff,\, QSO}}/\tau_{{\rm eff,\,
Ly\alpha}}$ as a function of the
ratio of the ionisation rate by the QSO and the background field,
$\omega=\Gamma_{\rm QSO}/\Gamma_{\rm UVB}$, 
i.e. the profile $\xi = (1+\omega/a)^{-0.5}$, where a strength parameter
$a$ is introduced. The strength parameter measures the deviation from
the theoretical background model and is used to quantify
any influence of the environmental density field.
We improve the statistical analysis of the profile fitting in employing
a moving average to the profile. We reproduce an unbiased measurement 
of the proximity effect which is not affected by the host halo mass. The scatter
between the different lines of sight and different quasar host positions
increases with decreasing redshift, $\sigma_{\log a} \approx 0.08, 0.20 \textrm{ and } 0.36$ for $z = 4.8,
4, \textrm{ and } 3$, respectively. Around the host halos, we find only a slight average 
overdensity in the proximity zone at comoving radii of $1<r_c<10 h^{-1} \textrm{ Mpc}$. However,
a clear power-law correlation of the strength parameter with the average overdensity in $r_c$ is found, 
showing an overestimation of the ionising background in overdense regions and an underestimation in
underdense regions.
\end{abstract}


\section{Introduction}

The Lyman alpha (Ly$\alpha$) forest in high-resolution quasar (QSO) spectra blueward of the 
QSO Ly$\alpha$ emission represents an
excellent tracer of the high redshift matter distribution. According to
detailed simulations and a large
observational material \citep[cp. e.g.][]{Meiksin:2009xq} the forest of lines
is generally ascribed to the absorption
by a tiny fraction of remaining neutral hydrogen \ion{H}{i}. This \ion{H}{i} forms
the cosmic web at low gas overdensities of a factor of a few and arises naturally due to
gravitational instabilities \citep{Petitjean:1995yq, Miralda-Escude:1996fv, Hui:1997fk}
in a standard $\Lambda$CDM cosmological model. The intergalactic gas
distribution and its ionisation state forms
from a balance of the ionising intergalactic UV radiation stemming from
galaxies and quasars and the
developing inhomogeneous density field. Due to the high ionisation
degree, the mean free path of UV photons in the Ly$\alpha$ forest is hundreds of Mpc
long. Hence, many sources which are distributed over large distances
contribute to the UV background (UVB), creating a quite
homogeneous UVB. 

This picture changes slightly in the vicinity of a QSO. 
There the QSO radiation dominates over the overall
background, additionally increasing the ionisation state of the intergalactic gas. This
manifests itself in a reduction of the number of observed absorption features towards the
emission redshift of the QSO.
This Òproximity effectÓ was first observed in low resolution spectra by \citet{Carswell:1982yt} and 
was later confirmed by \cite{Murdoch:1986ph} and \citet{Tytler:1987we}. 
In a seminal paper, \citet{Bajtlik:1988kx} used the proximity effect to estimate the
intensity of the intergalactic UV radiation in comparison to the QSO
luminosity, assuming that QSOs reside in random regions of the universe. 
Investigations of the proximity effect along these lines using absorption line
counting statistics \citep{Lu:1991rw, Williger:1994xe, Cristiani:1995zi, Giallongo:1996cs, 
Srianand:1996gd, Cooke:1997vn, Scott:2000ys}
or using pixel statistics of the transmitted flux 
\citep{Liske:2001uq, DallAglio:2008uq, DallAglio:2008ua, DallAglio:2009ud, Calverley:2010dp}
reveal the proximity effect in many high-resolution QSO spectra. 
Recent results by \citet{DallAglio:2008ua}, \citet{Calverley:2010dp}, and \citet{Haardt:2011jk} 
indicate a steady drop in the ionising background radiation's photoionisation
rate $\Gamma_\textrm{UVB} \sim 10^{-11.7} \textrm{ s}^{-1}$ at $z \sim 2$ 
to $\Gamma_\textrm{UVB} \sim 10^{-13.85} \textrm{ s}^{-1}$ for $z \gtrsim 5.5$.
However results obtained from SDSS spectra indicate a constant ionising background
level between $2.5 < z < 4.6$ \citep{DallAglio:2009ud}.

Independent measurements of the UV background radiation can be obtained with the flux
decrement method, which requires cosmological simulations to model statistical properties of
the Ly$\alpha$ forest \citep{Rauch:1997ms, Theuns:1998rw, Songaila:1999lj, McDonald:2001qe,
Meiksin:2003kh, Tytler:2004kz, Bolton:2005yq, Kirkman:2005vo, Jena:2005zx, Faucher-Giguere:2008ux,
Bolton:2007zf}.
It becomes clear that measurements using the proximity effect seem to overestimate the UV background
by a factor of a few. It is discussed whether this discrepancy arises from environmental effects 
such as clustering, gas infall, or the large scale density environment, which are neglected in the proximity 
effect modelling \citep{Rollinde:2005uq, Guimaraes:2007uq, Faucher-Giguere:2008gf,
Partl:2010qm}. Such effects can result in an overestimation of the UV background by up to 
a factor of 3 \citep{Loeb:1995kl}.

QSOs are thought to reside in very massive halos. Observational estimates on the QSO halo
mass revealed values of a couple of $10^{12} M_{\astrosun}$ \citep{da-Angela:2008ux} up
to $\sim 10^{14} M_{\astrosun}$ \citep{Rollinde:2005uq}. From numerical simulations of structure 
formations it is known that such massive halos are not located at random
positions in the universe, but form in dense environments where galaxies cluster. 
Observational determinations of the
large scale environment around QSOs from Ly$\alpha$ forest spectra indicate that QSOs
are embedded in large scale overdensities extending from proper $\sim 3-5 \textrm{ Mpc}$
\citep{DOdorico:2008az} to proper $\sim 10-15 \textrm{ Mpc}$ \citep{Rollinde:2005uq, Guimaraes:2007uq}. 
Numerical simulations by \citet{Faucher-Giguere:2008gf} show large scale 
overdensities of proper $\sim 3 - 6 \textrm{ Mpc}$ for redshifts $4 > z > 2$, consistent with
the results by \citet{DOdorico:2008az}. Using detailed radiative transfer simulations of three
QSOs residing in different cosmic environments \citet[][hereafter P1]{Partl:2010qm}
found indications that such large scale overdensities weaken the apparent proximity
effect signal, resulting in an overestimation of the UV background. 
However, the low number of QSO hosts in the P1 study
does not allow us to securely confirm the effect. We therefore extend this study using a large
sample of different dark matter halos in various mass ranges and inquire how such large scale
overdensities affect UV background measurements. It is also checked whether a possible 
dependence of the UV background with the host halo mass exists, as was suggested 
by \citet{Faucher-Giguere:2008gf}.

This paper is structured as follows. In Sect. \ref{sub:p2_simulations} we present the dark-matter
simulation used for this study and determine realistic models of the Ly$\alpha$ forest from
a semi-analytical model of the intergalactic medium. In Sect. \ref{sec:p2_method} we introduce
the proximity effect as a measure of the intergalactic ionising background flux and characterise
the halo sample used in this study. It is further discussed how Ly$\alpha$ forest mock spectra 
are generated. Subsequently in Sect. \ref{sec:p2_results}
we evaluate the effects of large scale overdensities and infall velocities
on the proximity effect. A possible dependence
of the UV background measurements on the halo mass and on the large scale mean density around the 
QSO are assessed. We summarise our results in Sect. \ref{sec:p2_conclusions}.

\section{Simulation}
\label{sub:p2_simulations}
\subsection{Distribution of baryons in the IGM}
\label{sub:modelIGM} 

In order to obtain a model for the gas content in the IGM that is in agreement with observed properties
of the Ly$\alpha$ forest, we employ the $64 \;  h^{-1} \textrm{ Mpc }$\footnote{Distances are given as comoving distances unless otherwise stated.}
DM simulation of the CLUES project\footnote{http://www.clues-project.org}
\citep{Gottloeber:2010xp} with $1024^{3}$ DM particles. 
The GADGET2 \citep{Springel:2005xz} simulation has a mass resolution of 
$m_{\mathrm{p, DM}}=1.86\times10^{7}\;M_{\astrosun} \; h^{-1}$ and uses a WMAP5 
\citep{Hinshaw:2009vn} cosmology.

The distribution of DM particles was obtained at four different redshifts
$z=4.8,\, 4,\, 3, \textrm{and }2$. Using triangular shaped cloud (TSC) 
density assignment we obtain a $800^3$ regularly spaced density and velocity 
grid from the DM particle distribution. Halos have been identified in the simulation using the
hierarchical friends-of-friends (HFOF) algorithm \citep{Klypin:1999bs} with a linking 
length of $0.17$.

\begin{figure*}
\includegraphics[width=0.48\textwidth]{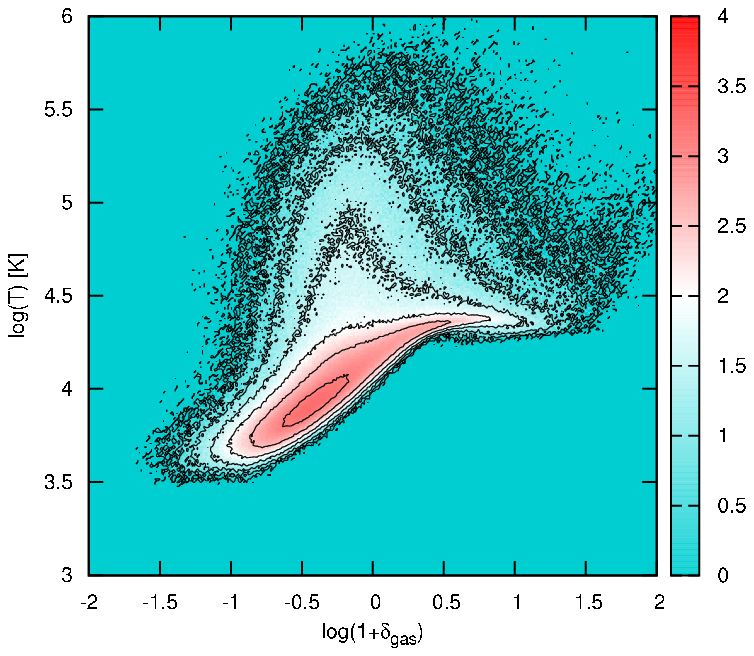}
\hfill{}%
\includegraphics[width=0.48\textwidth]{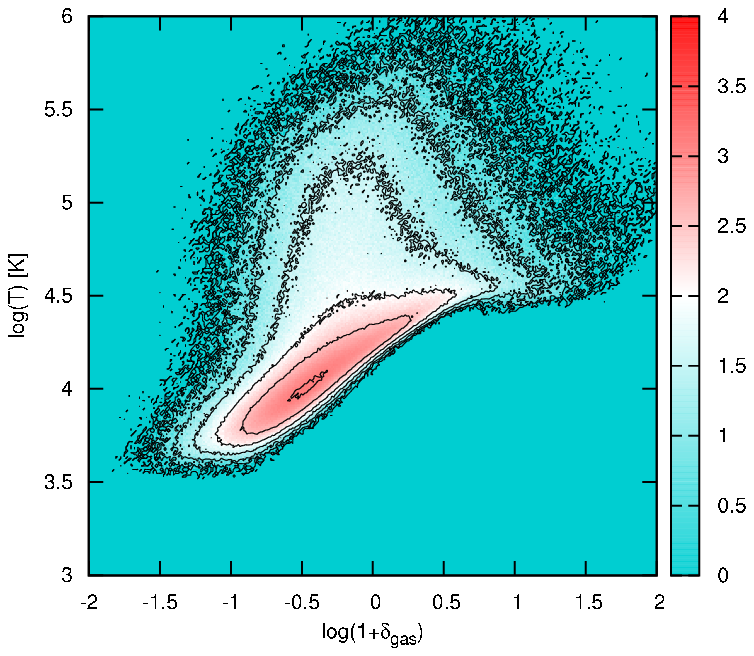}\\
\includegraphics[width=1\columnwidth]{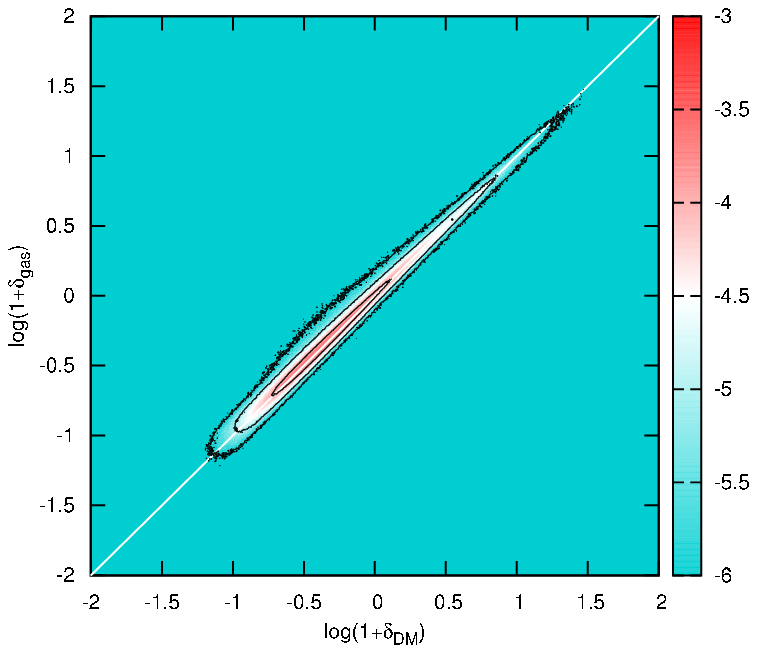}
\hfill{}%
\includegraphics[width=0.48\textwidth]{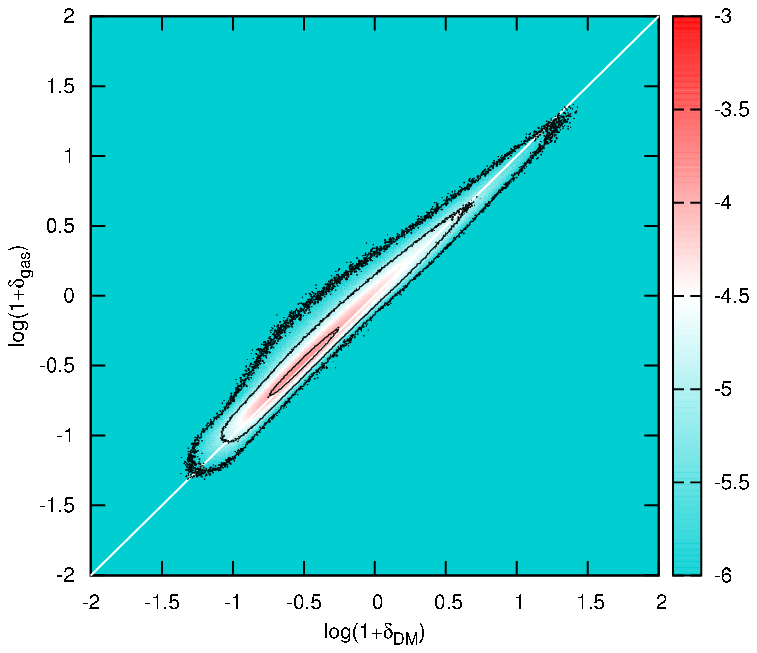}\\
\includegraphics[width=0.43\textwidth]{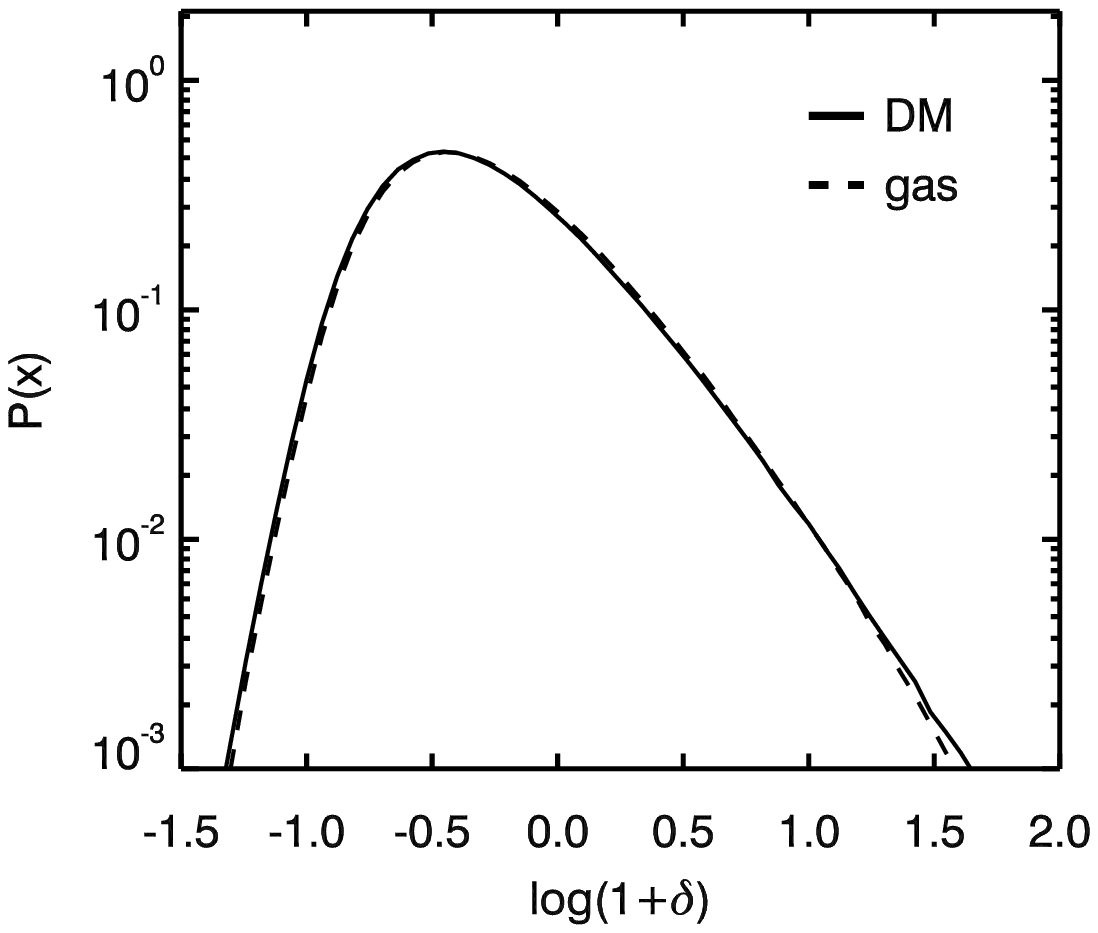}
\hfill{}%
\includegraphics[width=0.43\textwidth]{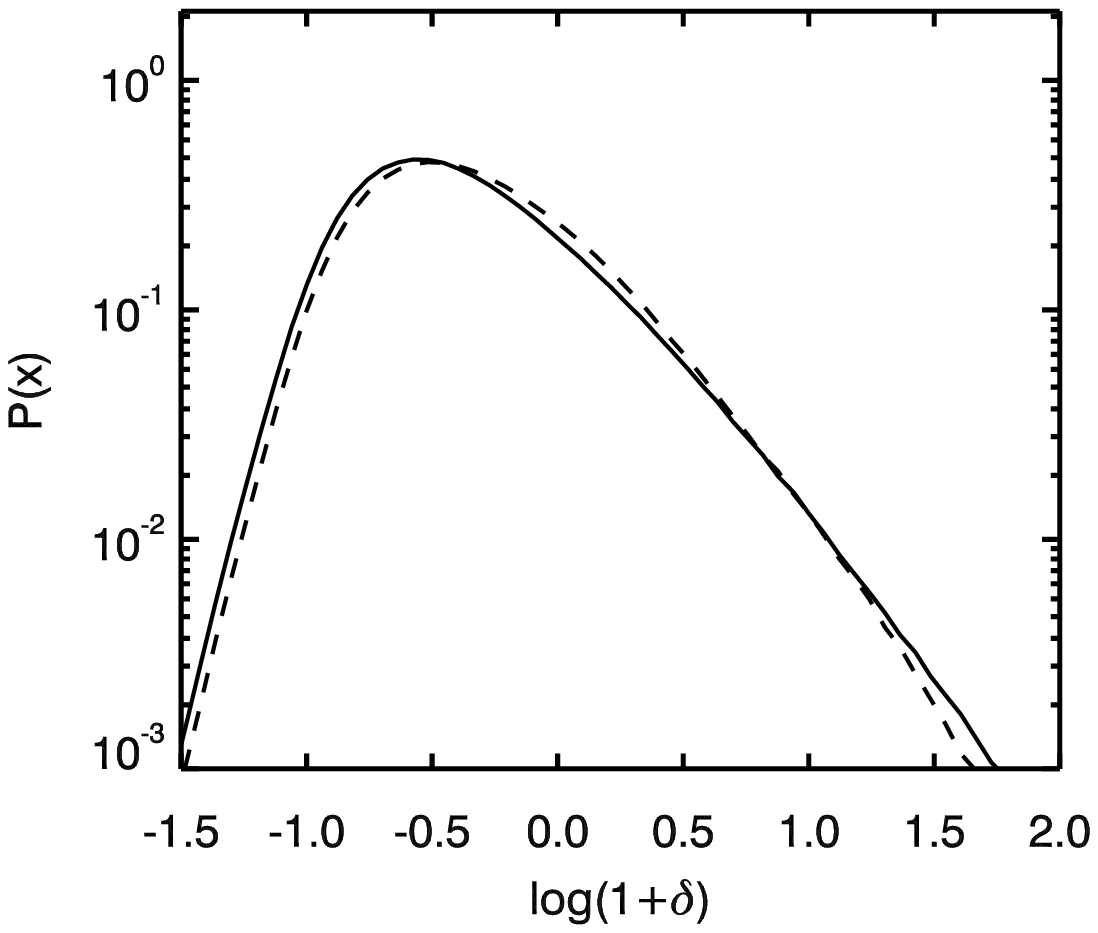}%
\hfill{} ~\\

\caption{\label{fig:12.5SimuRes} The top row shows the effective equation of state of a
$12.5 \;  h^{-1} \textrm{ Mpc}$ gasdynamic simulation binned to cells of $80 \;  h^{-1}  \textrm{ kpc}$
at redshift $z=5$ (left panels) and $z=4$ (right panels). The middle row shows the gas overdensity 
$1+\delta_{\mathrm{gas}}$ as a function of dark matter density $1+\delta_{\mathrm{DM}}$. The
white diagonal line marks the linear relation if gas strictly follows the dark matter. In both rows
the colour coding refers to the logarithm of the respective number of cells.  The lowest row
shows the probability distribution of the dark matter (solid line) and gas (dashed line) overdensity.}

\end{figure*}

To obtain an IGM gas density field from which we can construct \ion{H}{i} Ly$\alpha$ forest 
spectra, it is assumed that the properties of the baryonic component, such as density and bulk velocity,
are proportional to those of the dark matter \citep{Petitjean:1995yq, Meiksin:2001tz}. We have checked this assumption
with a comparable GADGET2 gas dynamical simulation using 
$2\times256^3$ particles with a size of $12.5 \;  h^{-1} \textrm{ Mpc}$ 
(Forero-Romero et al 2011, {\it in preparation}). This corresponds 
to a mass resolution of $m_{\mathrm{p, bar.}} = 3.7 \times 10^{5}\;M_{\astrosun}\;h^{-1}$ in gas. 
The gas dynamical SPH simulation includes also radiative and compton cooling, star formation, and
feedbacks through galactic winds using the model of \citet{Springel:2003rw}. Furthermore, a UVB 
generated from QSOs and AGNs is included at $z<6$ \citep{Haardt:1996qc}.
The state of the simulation has been recorded at $z=5 \textrm{ and } 4$. 
The simulation was rerun with DM only using the same realisation of the initial conditions. We obtain
density fields from the two simulations with equal spatial resolution as the larger
$64 \;  h^{-1} \textrm{ Mpc }$ sized fields. The DM densities are then compared to the densities derived 
from the gas dynamical simulation and the 
resulting DM to gas density relation is shown in Fig. \ref{fig:12.5SimuRes}. The baryonic component
follows the DM density closely, especially at $z=5$. For lower redshifts a linear relation is still
obtained, however the scatter around the line of equality, where the DM overdensity is equal to the gas overdensity, increases. A small bump in the DM to gas density relation towards higher gas densities
develops with decreasing redshift at gas densities where a fraction of the gas becomes shock heated.
The density in these shocks is slightly larger than the underlying DM, giving rise to the small
bump in the DM to gas density relation. Studying the density
probability distribution reveals that at $z=5$ the two components are similar (see Fig.
\ref{fig:12.5SimuRes}). At lower redshifts, the baryon distribution is slightly shifted to 
higher densities and decreases steeper than the DM one at high densities. For completeness we additionally
show the effective equation of state derived from the density and temperature fields in 
Fig. \ref{fig:12.5SimuRes}.

From the gas dynamical simulation we find the assumption that baryons follow the DM distribution to be reasonable
for the low density regions from which the Ly$\alpha$ forest arises. We therefore use the DM only simulations to derive 
the gas density and velocity fields as described in P1. By using a TSC mass assignment
scheme, we implicitly smooth our density field on sizes of 1.5 cells (i.e. 120 $h^{-1}$ kpc). 
This is comparable to a constant Jeans length smoothing of $\lambda_{\mathrm{Jeans}} \propto 
\delta^{-1/2}(1+z)^{-1/2} \sim 150 \, h^{-1} \textrm{ kpc}$ at $z=3$, where $\delta$ is the density contrast.

\subsection{Model and calibration of the intergalactic medium}
\label{sub:calib}

The characteristics of the IGM gas are determined analogously to the method
described in P1. We will therefore only briefly sketch the method with which a 
representative IGM model is obtained. 

The temperature in the IGM can be approximated by the so called 
effective equation of state \citep{Hui:1997uq}, which is
expressed as $T=T_{{\rm 0}}(1+\delta)^{\gamma-1}$. The effective equation of state shows
a linear trend for overdensities $0.1 \lesssim \delta \lesssim 10$ (compare Fig. \ref{fig:12.5SimuRes}). 
For higher densities its behaviour is approximated by assuming a constant value above 
a temperature cut-off $T_\mathrm{cut-off} = T\left( \delta = 10 \right)$. 
By employing the method developed in \citet{Hui:1997fk}, we compute \ion{H}{i} Ly$\alpha$ 
absorption spectra from the DM 
density and velocity fields, once the parameters $T_{{\rm 0}}$, $\gamma$, and the UV background 
photoionisation rate ($\Gamma_{\rm UVB}$) are determined. These are constrained by matching
statistical properties of the simulated \ion{H}{i} absorption spectra with observed relations.
Such a calibration will be important in obtaining realistic representations of the IGM that
closely mimic the observed characteristics.

In order to calibrate our IGM model, we employ four observational constraints sorted by
increasing importance: (i) The observed equation of state \citep{Ricotti:2000nx,
Schaye:2000kx, Lidz:2010oz}, (ii) the evolution of the UV background photoionisation rate 
\citep{Haardt:2001yq,Bianchi:2001ys,Bolton:2005yq,DallAglio:2008uq, DallAglio:2009ud}, 
(iii) the observed evolution of the effective 
optical depth in the Ly$\alpha$ forest \citep{Schaye:2003kx, Kim:2007kx}, and 
(iv) the transmitted flux probability distribution \citep[FPD,][]{Becker:2006uq}.

The statistical quantities of the simulated spectra are derived from the simulation using
500 lines of sight randomly drawn through the cosmological box. Due to the availability of
high resolution spectra with high signal-to-noise (S/N) ratios of up to 120, we approximate such
high quality data by assuming noise free spectra. It has been noted by \citet{Calverley:2010dp}
that low S/N levels introduce a systematic shift in the proximity effect signal. Since we want
to quantify the physical effect of the halo's surrounding environment on UVB measurements, adding 
noise would only lead to degeneracies between the two effects. However we will briefly address the 
influence of noise in Section \ref{sec:proxySN}.

The spectra are convolved with the instrument profile of the UVES spectrograph
and are then binned to the typical resolution of UVES spectra of 
$6.7 \textrm{ km s}^{-1}$. We further assume that the QSO continuum can be perfectly determined and we 
therefore consider no uncertainties in the continuum. This procedure is adopted to precisely 
quantify physical effects without contamination of the signal with observational uncertainties. These will 
only increase the variance in the results discussed below.

\begin{figure}
\includegraphics[bb=0bp 0bp 360bp 278bp,clip,width=1\columnwidth]{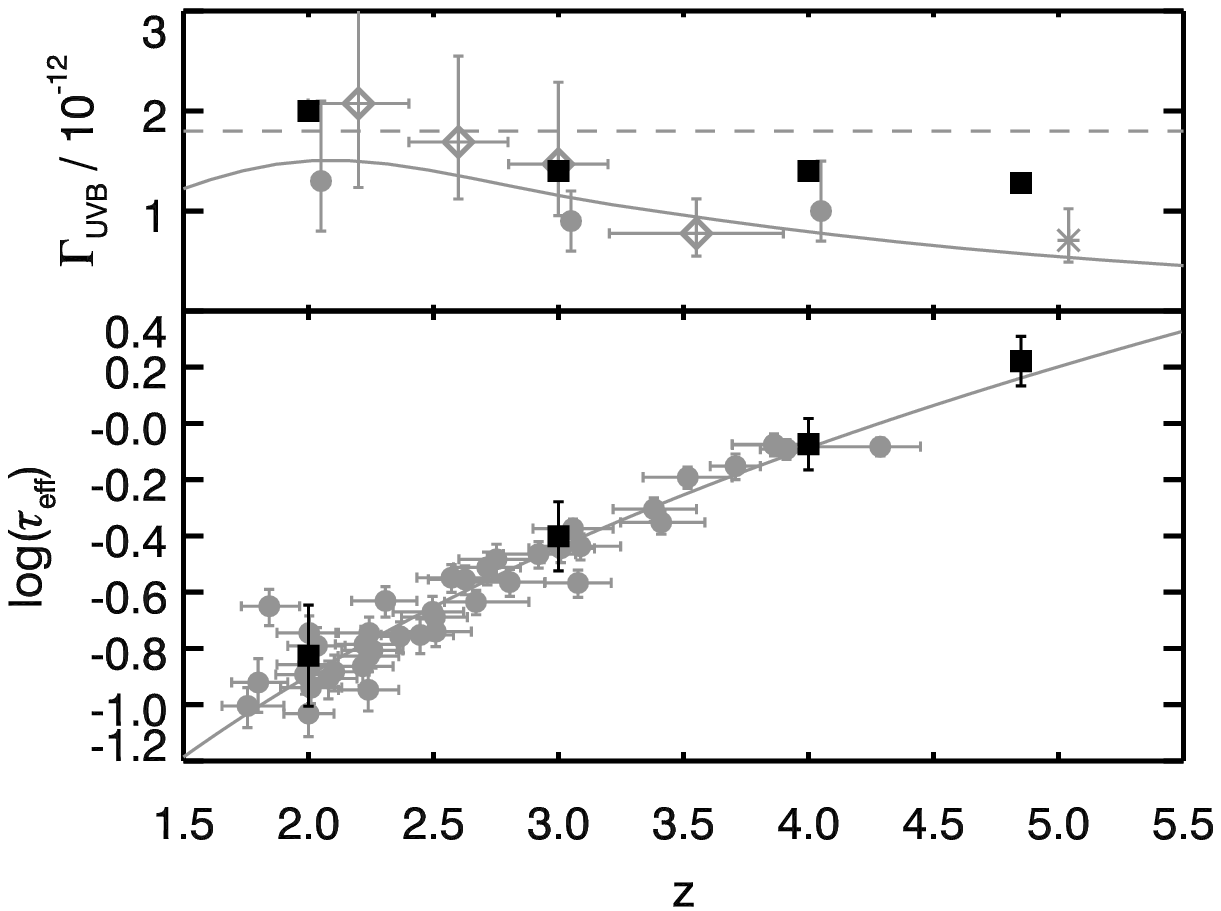}

\caption{\label{fig:meanTauEff}\emph{Upper panel: } The evolution of the UV background
photoionisation rate in the four snapshots (black squares) compared to \citet{Bolton:2005yq} (grey points
shifted by $\Delta z=0.05$ for better visibility), measurements by \citet{DallAglio:2008ua} 
(grey open diamonds) and
\citet{Calverley:2010dp} (grey star), a fit to SDSS data measurements by 
\citet{DallAglio:2009ud} (grey dashed line)
and predictions by \citet{Haardt:2001yq} (grey line).
\emph{Lower panel: }The effective optical depth of our models (black
points) in comparison to measurements by \citet{Schaye:2003kx} (grey
points). The continuous line shows the fit to observational data by \citet{Kim:2007kx}.}

\end{figure}

\begin{figure}
\includegraphics[width=1\columnwidth]{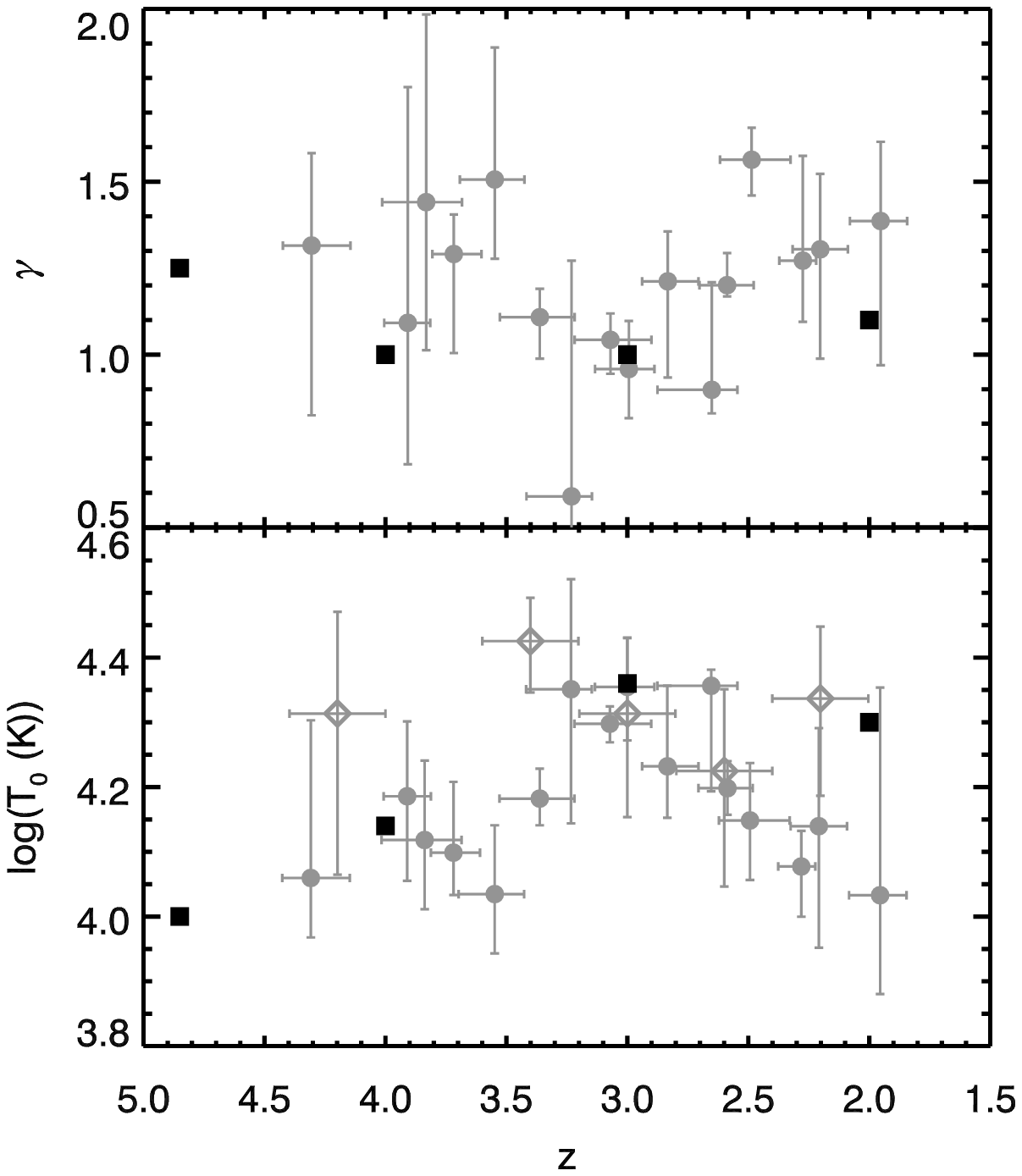}

\caption{\label{fig:EOS}\emph{Upper panel: } Comparison of our choices for $\gamma$ (black
squares) with observationally derived results by \citet{Schaye:2000kx} (grey points). 
\emph{Lower panel: }Comparison of our model $T_{\rm 0}$ (black points) with observations by
\citet{Schaye:2000kx} (grey points) and \citet{Lidz:2010oz} (grey open diamonds).}

\end{figure}

\begin{figure*}
\centering
\includegraphics[width=0.42\textwidth]{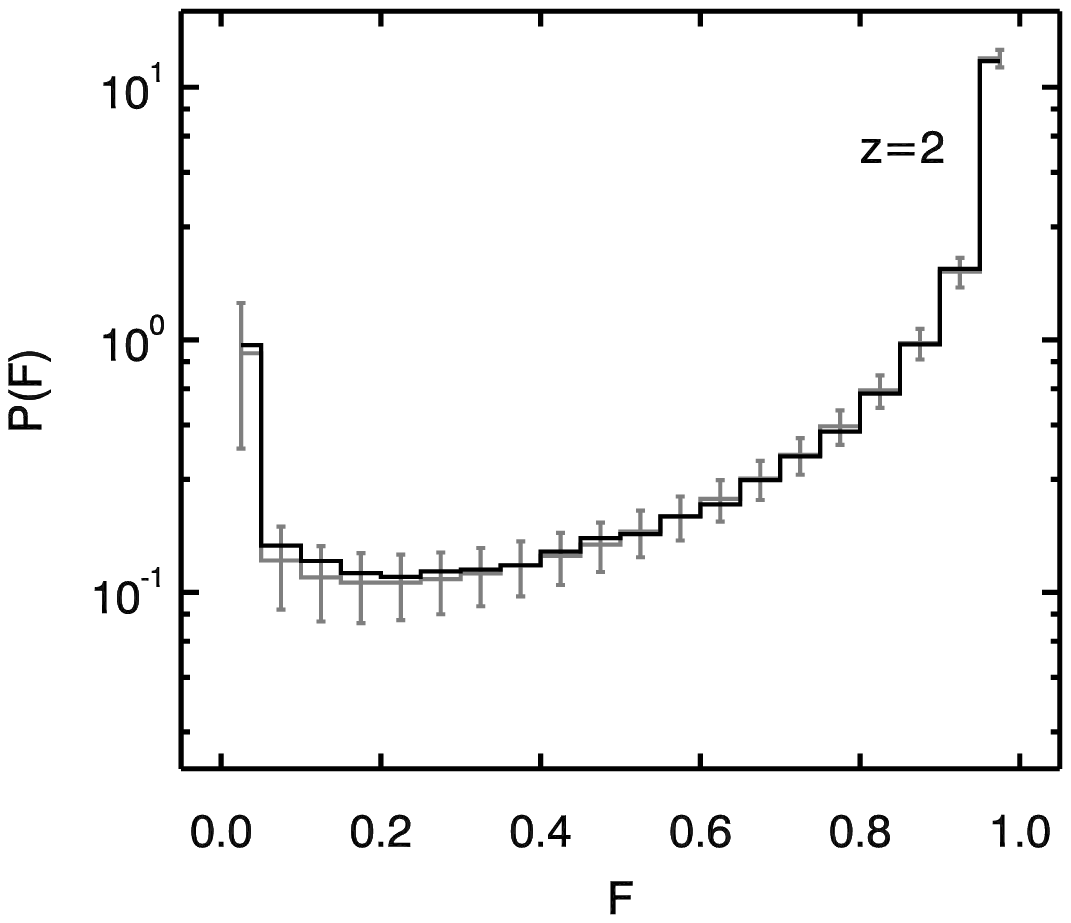}
\includegraphics[width=0.42\textwidth]{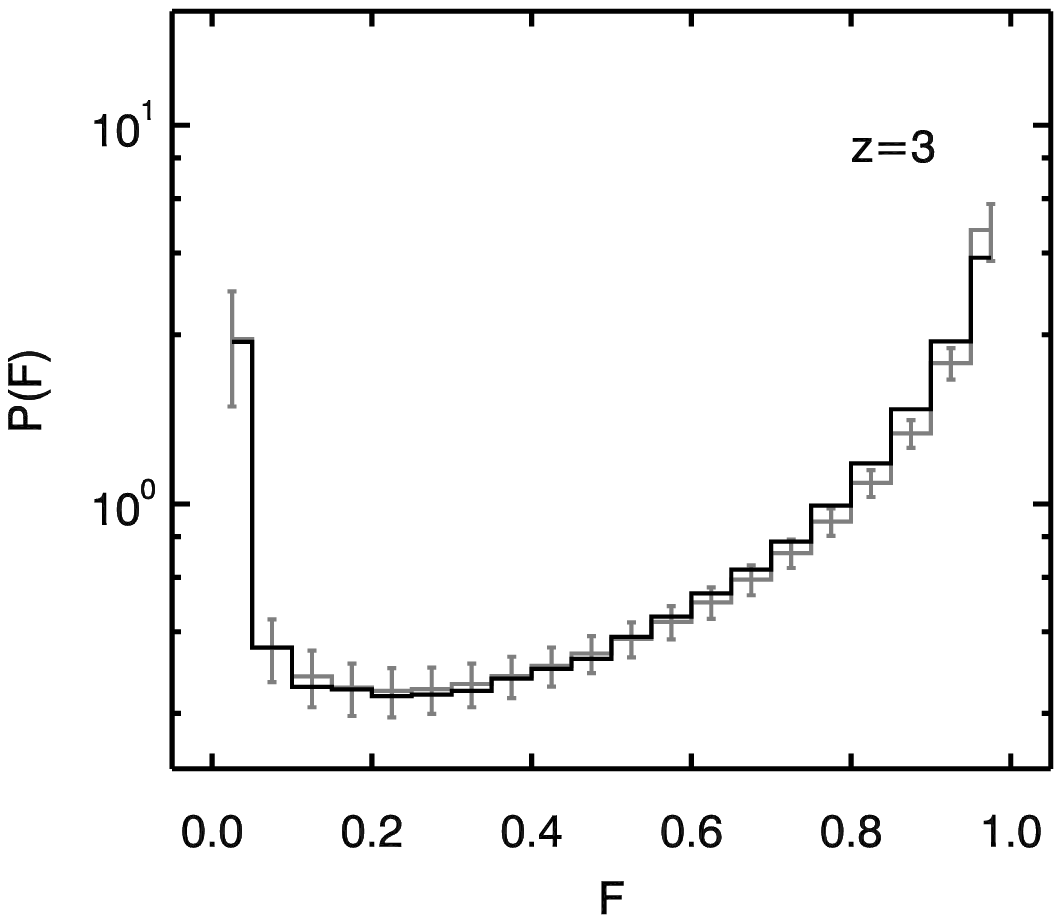}\\
\includegraphics[width=0.42\textwidth]{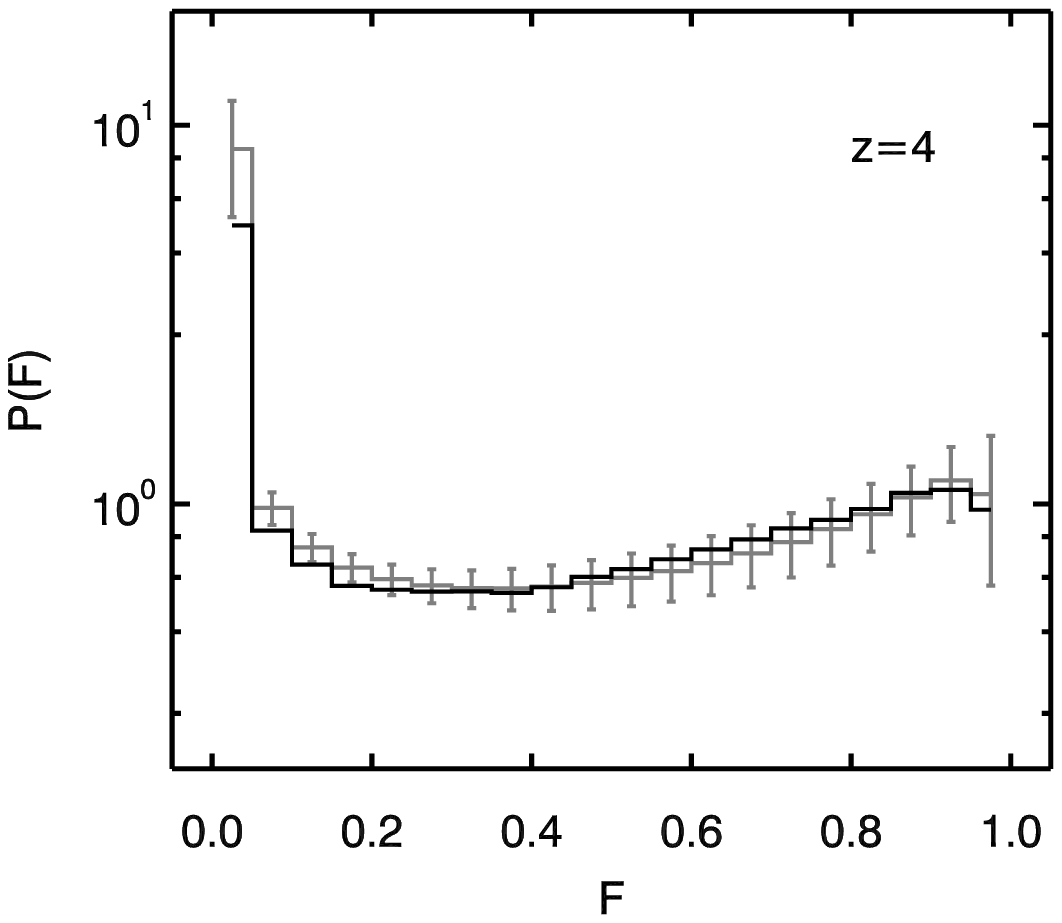}
\includegraphics[width=0.42\textwidth]{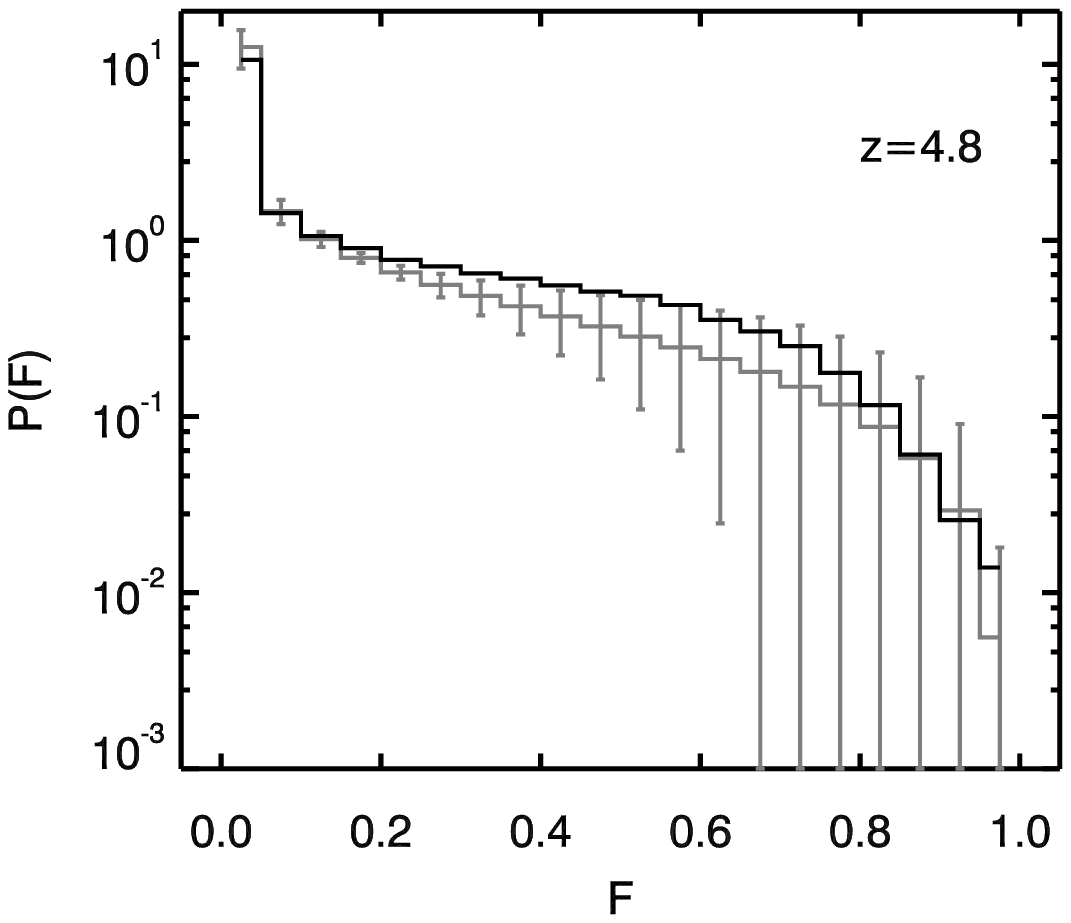}\\

\caption{\label{fig:PDFs}Probability distribution function of transmitted flux $F$ at redshifts 2, 3,
4, and 4.8. Grey lines give the mean observed PDF obtained from the 
fits excluding noise given in \citep{Becker:2006uq} with $1\sigma$ error bars denoting the variation 
between different lines of sight. The mean is derived in redshifts bins of $\pm 0.25$ centred on the model
redshift. The black full line shows the results from our models.}

\end{figure*}

\begin{figure*}
\centering
\includegraphics[width=0.42\textwidth]{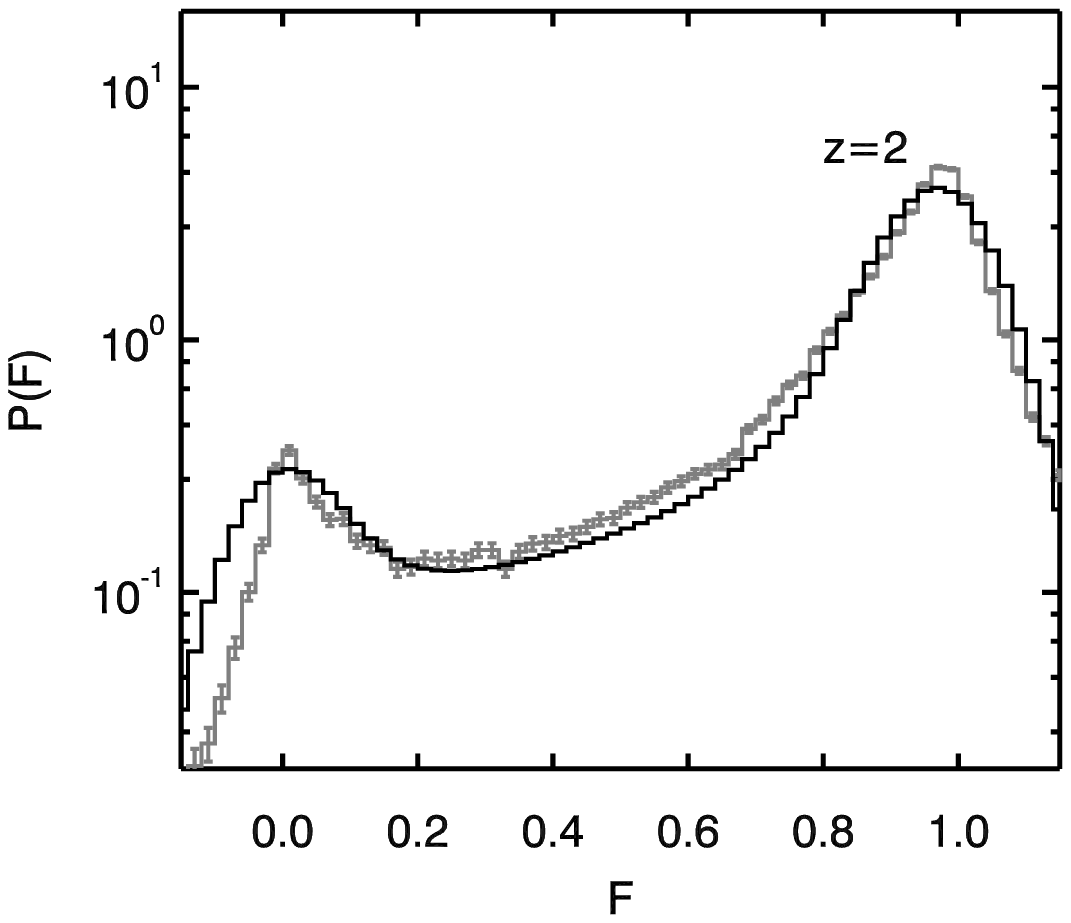}
\includegraphics[width=0.42\textwidth]{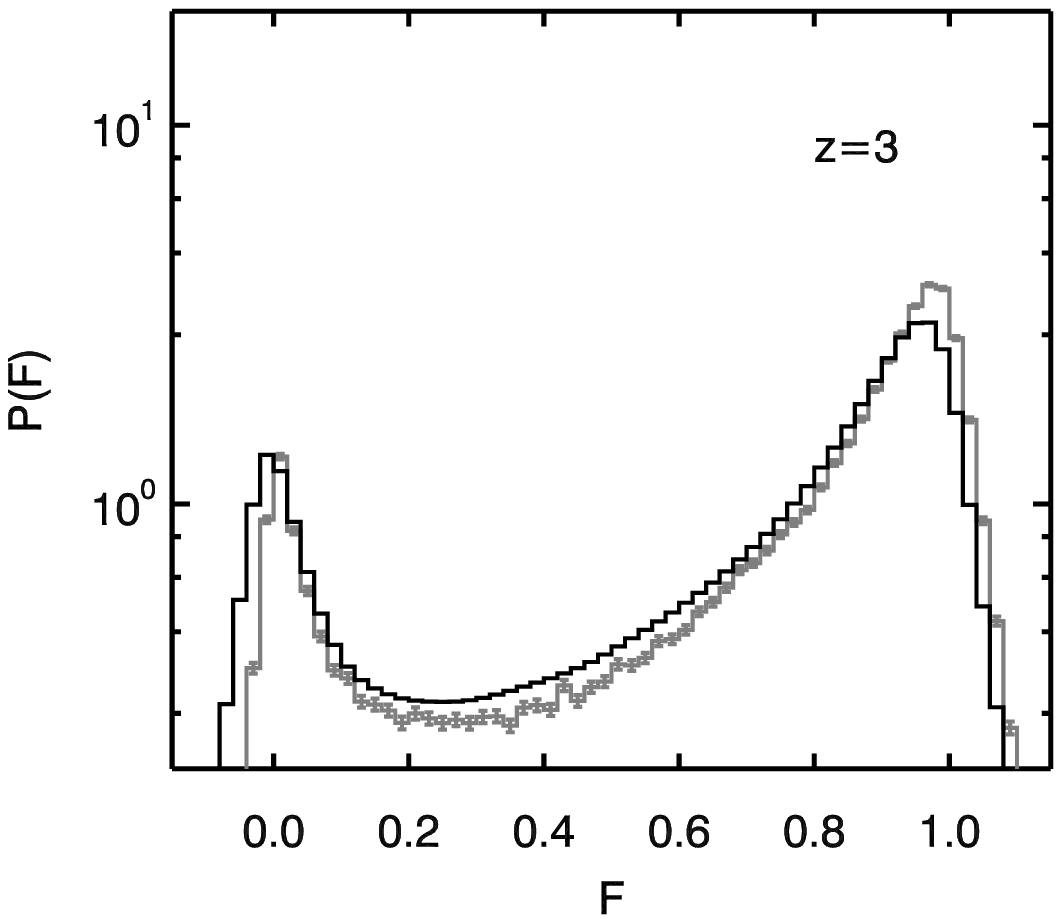}\\
\includegraphics[width=0.42\textwidth]{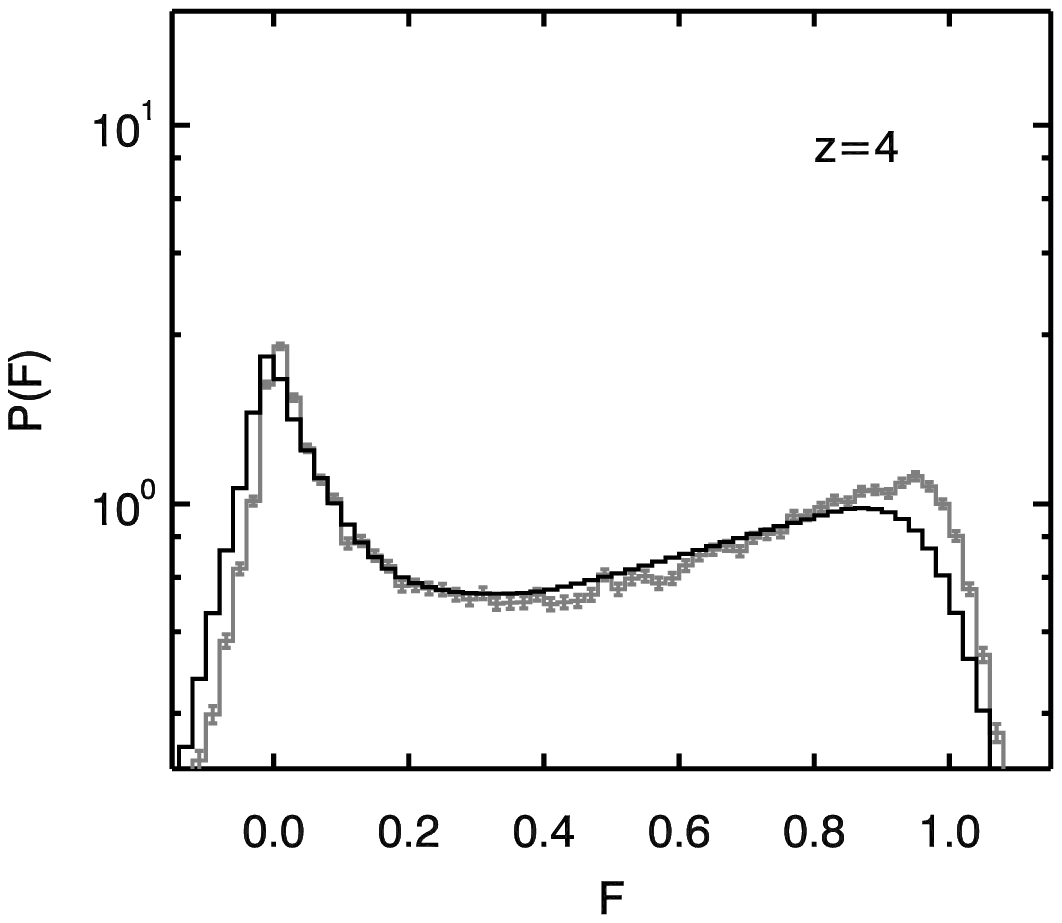}
\includegraphics[width=0.42\textwidth]{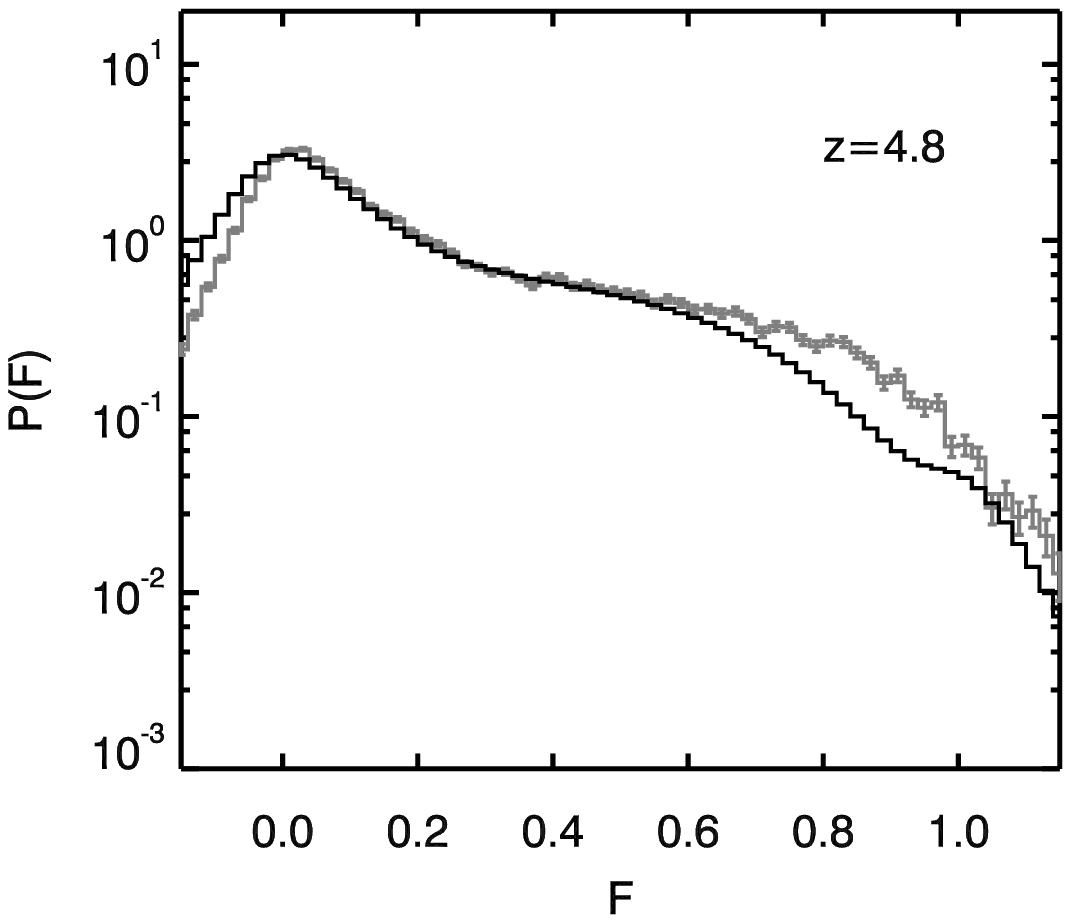}\\

\caption{\label{fig:PDFObss}Probability distribution function of transmitted flux $F$ at redshifts 2, 3,
4, and 4.8. Grey lines with $1\sigma$ error bars give the 
observed PDF \citep{Becker:2006uq} and the black full line show the results obtained from 
our models folded with the signal to noise function of the combined observations.
The combined observational data is derived in redshifts bins of $\pm 0.25$ centred on the model
redshift.}

\end{figure*}

To determine the model parameters for the IGM, we adopt an iterative method based on 
$\chi^2$ minimisation between simulated
and observed FPD, as the latter provides one of the strongest constrains on the 
Ly$\alpha$ forest properties. Our steps are as follows:

\begin{enumerate}
\item{
We choose a initial guess $\left(T_{{\rm 0}}, \gamma, \Gamma_{\rm
UVB}\right)_0$ according to the measurements of the equation of state by \citet{Schaye:2000kx} and
\citet{Lidz:2010oz} and the evolution of the UVB by \citet{Haardt:2001yq} (see Figs.
\ref{fig:meanTauEff} and \ref{fig:EOS}). }

\item{
With the above initial guess and using 500 lines of sight, we determine
the average effective optical depth $\tau_{\rm{eff}}(z)
=-\ln\left\langle F(z)\right\rangle $ with $F$ being the transmitted flux 
and the averaging is performed over the whole line of sight. As $T_{{\rm 0}}$ and 
$\gamma$ have a subdominant effect on $\tau_{\rm{eff}}(z)$, we
first tune $\Gamma_{\rm UVB}$ to obtain a match in the effective optical depths of the 
model spectra with 
recent observations from \citet{Kim:2007kx}. This results in a new set of
parameters $\left(T_{{\rm 0}}, \gamma, \Gamma_{\rm UVB}\right)_1$. }

\item{
Finally we construct our simulated FPD. As observational constraint we
employ the log normal fits obtained by \citet{Becker:2006uq} in a redshift 
interval of $\Delta z=\pm 0.25$ centred on the snapshot redshift $z$ (see Fig. \ref{fig:PDFs}). 
By using their fits to the FPD, our model spectra can be compared with the observations without 
considering the effect of detector noise. However for consistency
we also compare the simulated FPD with the raw combined
observational results by \citet{Becker:2006uq}, folding the simulation result with the global noise
function of the combined observed sample (see Fig. \ref{fig:PDFObss}).
The best fit parameters $\left(T_{{\rm 0}}, \gamma, \Gamma_{\rm
UVB}\right)_\star$ are then iteratively determined with a Simplex optimisation procedure. }
\end{enumerate}

Our best fit parameters are presented in Table \ref{tab:gnedModels}, and
plotted in Figs.~\ref{fig:meanTauEff}, \ref{fig:EOS}, \ref{fig:PDFs}, and \ref{fig:PDFObss} 
in comparison with different literature results. 

\begin{table}
\centering
\caption{\label{tab:gnedModels}Model parameters of the semi-analytical model.}

\begin{tabular}{cccccc}
\hline 
\hline$z$ & $\log(T_{0} [K])$ & $\gamma$ & $\Gamma_{\rm{UVB}} [\textrm{s}^{-1}]$ 
& $\log \tau_\mathrm{eff} $\\
\hline
$4.8$ & $4.00$ & $1.25$ & $1.16\times10^{-12}$ & $0.222$\\
$4.0$ & $4.14$ & $1.00$ & $1.4\times10^{-12}$ & $-0.074$\\
$3.0$ & $4.36$ & $1.00$ & $1.4\times10^{-12}$ & $-0.401$\\
$2.0$ & $4.30$ & $1.10$ & $2.0\times10^{-12}$ & $-0.826$\\
\hline
\end{tabular}
\end{table}

The inferred evolution of the UV background in Fig. \ref{fig:meanTauEff} closely follows recent results
by \citet{Haardt:2001yq}, \citet{Bolton:2005yq}, and \citet{DallAglio:2008uq}. The values for redshift 
$z=4.8$ is rather high when compared with \citet{Calverley:2010dp}, however it is still in agreement 
within the $2\sigma$ limits. Furthermore
the effective optical depth, also shown in Fig. \ref{fig:meanTauEff}, is consistent with high resolution observations 
by \citet{Schaye:2003kx} and \citet{Kim:2007kx}.
Additionally the final parameters of the equation of state $T_0$ and $\gamma$ lie within the 
measurement uncertainties of \citet{Schaye:2000kx} and \citet{Lidz:2010oz}, which is evident from
Fig. \ref{fig:EOS}.

The flux probability distributions estimated from the simulated sight lines agree
reasonably well with the noise-corrected observed profiles estimated by \citet{Becker:2006uq} who assumed
a log-normal distribution of the optical depth in the Ly$\alpha$ forest (see Fig. \ref{fig:PDFs}). 
At redshifts $z=2 \textrm{ and } 3$ the match between the mean observed profiles and our simulations is
good. However at higher redshifts, the agreement marginally decreases, even though the distributions are
consistent within the variation between different lines of sight. This discrepancy manifests itself
clearer when comparing the simulations with the raw observational data in Fig. \ref{fig:PDFObss} 
which include the effects of noise. The discrepancy is especially strong at the high transmission 
end of the distribution. This is most certainly caused
by our crude noise modelling of a sample with inhomogeneous S/N, which we obtained by
stacking the noise functions of the various lines of sight.  On the other hand
we assume a perfect knowledge of the continuum level in the spectra, which is 
a challenge to determine in observed spectra, especially at high redshifts.

\section{Method}
\label{sec:p2_method}
 
\subsection{Line-of-sight proximity effect}

Bright UV sources, such as QSOs, alter the ionisation state in their vicinity strongly up to
proper distances of a couple of Mpc. The resulting change in the ionisation state directly 
translates into a decrease in 
optical depth. This decrease manifests itself in a reduction of the absorption line density in the Ly$\alpha$
forest when approaching the redshift of the source. Using line-counting statistics, \citet{Bajtlik:1988kx}
measured the line-of-sight proximity effect in QSO spectra for the first time. They assumed 
that QSOs are situated in a mean IGM environment, neglecting any redshift distortions from peculiar 
velocities of the absorption clouds. They further assumed that the QSO's radiation field is radially 
decreasing from the source proportional to $r^{-2}$, i.e. geometric dilution. Using detailed radiative transfer 
simulations, P1 showed that radiative transfer effects only play a marginal role in the line-of-sight 
proximity effect and that the assumption of geometric dilution holds.

Assuming photoionisation equilibrium, the change in the optical depth as a function of distance $r$ to the
QSO can be expressed as
\begin{equation}
\label{eq:tauPE}
\tau_\mathrm{PE}(r) = \tau_\mathrm{Ly\alpha}(r) \left( 1 + \omega(r) \right)^{-1}
\end{equation}
\citep{Liske:2001uq}, where $\tau_\mathrm{PE}$ is the observed optical depth, $\tau_\mathrm{Ly\alpha}$
is the optical depth in the absence of a QSO, and
\begin{equation}
\label{eq:omegaFunc}
\omega(r) = \frac{\Gamma_\mathrm{QSO}(r)}{\Gamma_\mathrm{UVB}} = \frac{1}{16Ê\pi^2} \frac{1}{r^2} \frac{L_{\nu_\mathrm{LL}}}{J_{\nu_\mathrm{LL}, \mathrm{UVB}}} \frac{3-\alpha_\mathrm{b}}{3-\alpha_\mathrm{q}}
\end{equation}
acts as a normalised distance to the QSO \citep{DallAglio:2008uq}. 
Here $\Gamma_\mathrm{QSO}(r)$ and $\Gamma_\mathrm{UVB}$ are the photoionisation rates of the 
QSO and the UV background respectively, $L_{\nu_\mathrm{LL}}$ is the QSO's luminosity at the Lyman
limit, while $J_{\nu_\mathrm{LL}, \mathrm{UVB}}$ is the UVB's Lyman limit flux, $\alpha_\mathrm{b}$ is
the slope of the UVB's spectral energy distribution $\propto \nu^{\alpha_\mathrm{b}}$, and $\alpha_\mathrm{q}$
denotes the spectral slope of the QSO's emission $\propto \nu^{\alpha_\mathrm{q}}$. In this work we
assume for simplicity the UVB to be dominated by QSOs. We therefore follow \citet{Haardt:1996qc}
and chose the slope of the UVB to be $\alpha_\mathrm{b} = -1.5$. This is consistent with 
observations by \citet{Telfer:2002yq} who find $\alpha_\mathrm{q} = -1.57\pm0.17$. The effect of differing spectral
shapes between the UVB and the QSO has been studied in \citet{DallAglio:2008uq} and P1. 
The proximity effect is introduced into lines of sight drawn from our simulation boxes by
modifying the neutral hydrogen fraction in real space using $n_\mathrm{\ion{H}{i}, PE}(r) = 
n_\mathrm{\ion{H}{i}, Ly\alpha}(r) \left( 1 + \omega(r) \right)^{-1}$.

\subsection{Environments}

To study the impact of the large scale environment on the proximity effect measurement as a function
of QSO host halo mass, we pick halos in given mass ranges from the simulation to serve as QSO hosts. Around
each halo, 100 lines of sight with a length of two comoving box sizes ($128 \;  h^{-1} \textrm{ Mpc}$)
are randomly drawn from the box assuming periodic boundary conditions. 
Our sample of QSO host halos cover a
wide range of masses. At each redshift we use the 20 most massive halos, 50 halos with a mass
around $10^{12} M_{\astrosun}$, and 50 halos with a mass around $10^{11} M_{\astrosun}$. For the
$10^{12}$ and $10^{11} M_{\astrosun}$ mass objects we choose 50 halos from the mass sorted halos
with a mass larger than the cutoff. For the more massive bin, less than 50 halos with mass higher than 
the cutoff are present in the simulation at $z\geq4$. In this case we choose all the halos above $10^{12}
M_{\astrosun}$ and extend the range to lower masses until the sample consists of 50 halos. 
For these redshifts, the $10^{12} M_{\astrosun}$
mass range overlaps with the 20 most massive halos.
The covered mass intervals are given in Table \ref{tab:haloMassBins} as a function of redshift.
We further note that the halos in the $10^{11} M_{\astrosun}$ mass bin are certainly not massive enough
to host QSOs, since \citet{da-Angela:2008ux} for instance
estimated a QSO host halo mass of around $3 \times 10^{12}
M_{\astrosun}$ independent of redshift and luminosity from the 2dF-SDSS survey. However to establish any
dependency of the proximity effect signal with the host's mass, they are included for the purpose of an 
extreme low mass range.

\begin{table}
\centering
\caption{\label{tab:haloMassBins}Mass range of the halo samples as a function of redshift. The
$10^{12} \; M_{\astrosun}$ and $10^{11} \; M_{\astrosun}$ samples contain 50 halos each. The
20 most massive halos are given in units of $10^{12} \; M_{\astrosun}$, the other mass bins in
units corresponding to the mass bin.}

\begin{tabular}{cccc}
\hline 
\hline$z$ & 20 most massive & $10^{12} \; M_{\astrosun}$ & $10^{11} \; M_{\astrosun}$\\
\hline
$4.8$ & $2.07 - 0.92$ & $2.07 - 0.53$ & $1.04 - 1.00$ \\
$4$ & $4.84 - 1.51$ & $4.84 - 0.88$ & $1.03 - 1.00$ \\
$3$ & $15.8 - 3.41$ & $1.38 - 1.00$ & $1.01 - 1.00$ \\
$2$ & $40.2 - 6.19$ & $1.14 - 1.00$ & $1.01 - 1.00$ \\
\hline
\end{tabular}
\end{table}

In order to test the influence of the large scale environment around QSO host halos we additionally construct
a sample of lines of sight for each redshift where the origins
are randomly selected. This sample
provides a null hypothesis since any influence of large scale density fluctuations 
averages out if the origins of the lines of sight are randomly distributed in the box.

\subsection{Measuring the proximity effect in spectra }
\label{sec:newProxMeth}

The proximity effect signature is measured in Ly$\alpha$ forest spectra by first constructing
an appropriate $\omega$-scale for the observed QSO. In this study we assume the QSO to have a
Lyman limit luminosity of $L_{\nu_\mathrm{LL}} = 10^{31} \textrm{ erg Hz}^{-1} \textrm{ s}^{-1}$, and 
the photoionisation rates of our model UVB field are used. We have
shown in P1 that for higher QSO luminosities the proximity effect is more pronounced and less
affected by the density distribution. Therefore we use 
a low luminosity QSO to obtain upper limits of the environmental bias.
Given the $\omega$-scale, the transmission
spectra is then evaluated for each line of sight in bins of $\Delta \log \omega$ \citep{DallAglio:2008uq}. 
Using the mean transmission per bin, the effective optical depth in the bin 
$\tau_{\mathrm{eff, QSO}}(\Delta \log \omega)$ 
is calculated and normalised to the effective optical depth in the Ly$\alpha$ forest unaffected by the QSO's
radiation $\tau_{\mathrm{eff, Ly\alpha}}$. The normalised optical depth $\xi$ is thus
\begin{equation}
\label{eq:P2normOptDepth}
\xi(\Delta \log \omega) = \frac{\tau_{\mathrm{eff, QSO}}(\Delta \log \omega)}{\tau_{\mathrm{eff, Ly\alpha}}}.
\end{equation}
The imprint of the proximity effect onto the normalised optical depth for a given $\omega$-scale becomes
\begin{equation}
\label{eq:P2xiOmega1}
\xi(\omega) = \left( 1+\omega(r) \right)^{1-\beta}
\end{equation}
\citep{Liske:2001uq} where $\beta$ is the slope of the Ly$\alpha$ absorber's column density distribution.
Throughout this work we assume $\beta = 1.5$ \citep{Kim:2001fk}. 
The UVB photoionisation rate $\Gamma_\mathrm{UVB}$ 
can then be determined using the proximity
effect strength parameter $a$
\begin{equation}
\label{eq:P2strengthParam}
\xi = \left( 1+ \frac{\omega}{a} \right)^{1-\beta}.
\end{equation}
This parametrisation was introduced by \citep{DallAglio:2008uq, DallAglio:2008ua} 
in analyses of observed spectra with an assumed UVB as reference.
Values of $a > 1$ or $a < 1$ indicate a weaker or stronger
proximity effect than the model, respectively. The measured photoionisation rate of the UVB is then
determined using the reference value $\Gamma_\mathrm{UVB, ref}$ multiplied by the strength
parameter $a$. 
In our case the strength parameter $a$ indicates any deviation of the measured UVB photoionisation
rate from the input value. The strength parameter is determined by fitting Eq. \ref{eq:P2strengthParam}
to the binned normalised optical depth $\xi(\Delta \log \omega)$. In order to exclude any
direct impact of the host halo on the strength parameter fit, only data with $\log \omega < 2$ have been 
used.

\begin{figure*}
\includegraphics[width=0.48\textwidth]{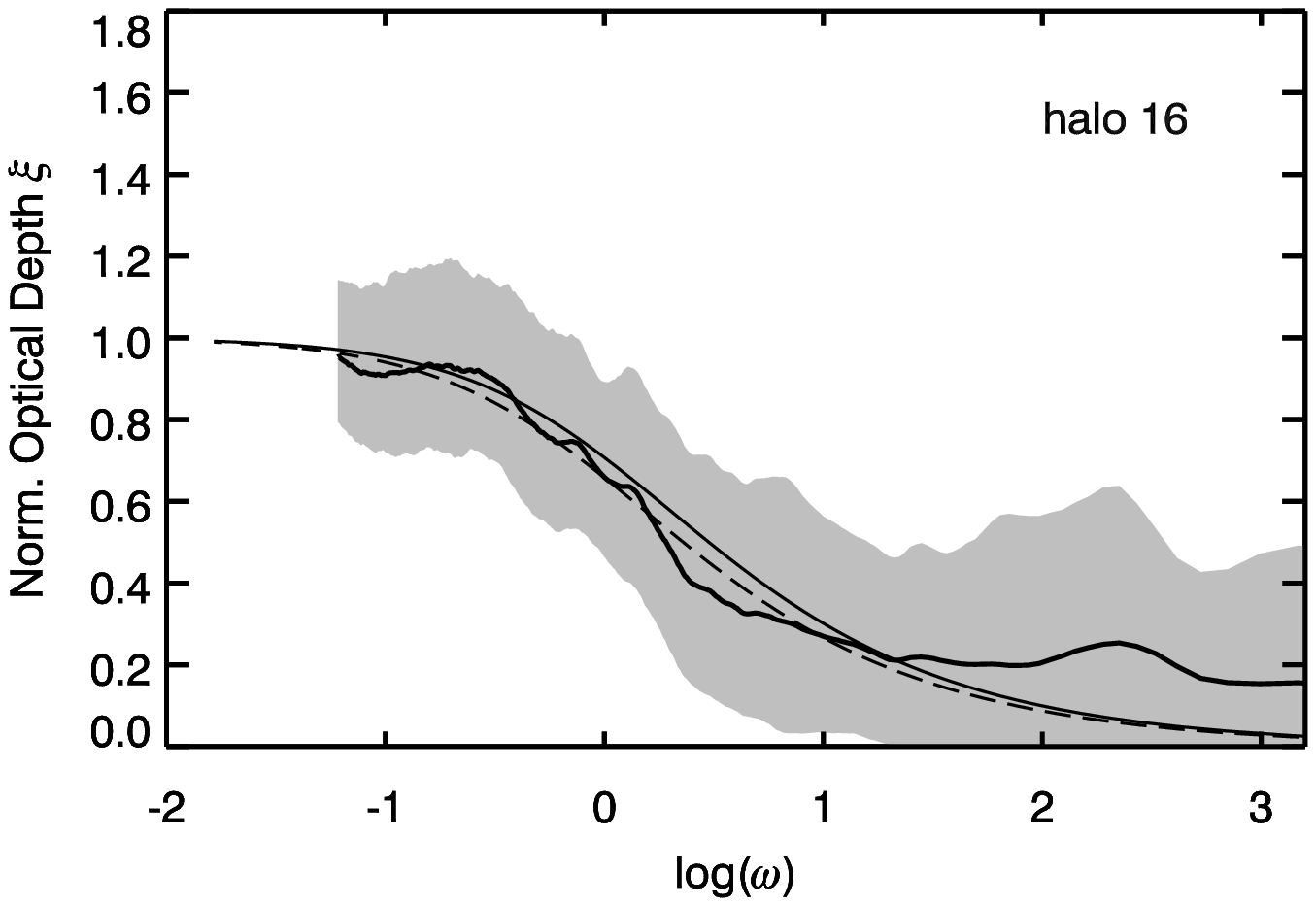}
\hfill{}%
\includegraphics[width=0.48\textwidth]{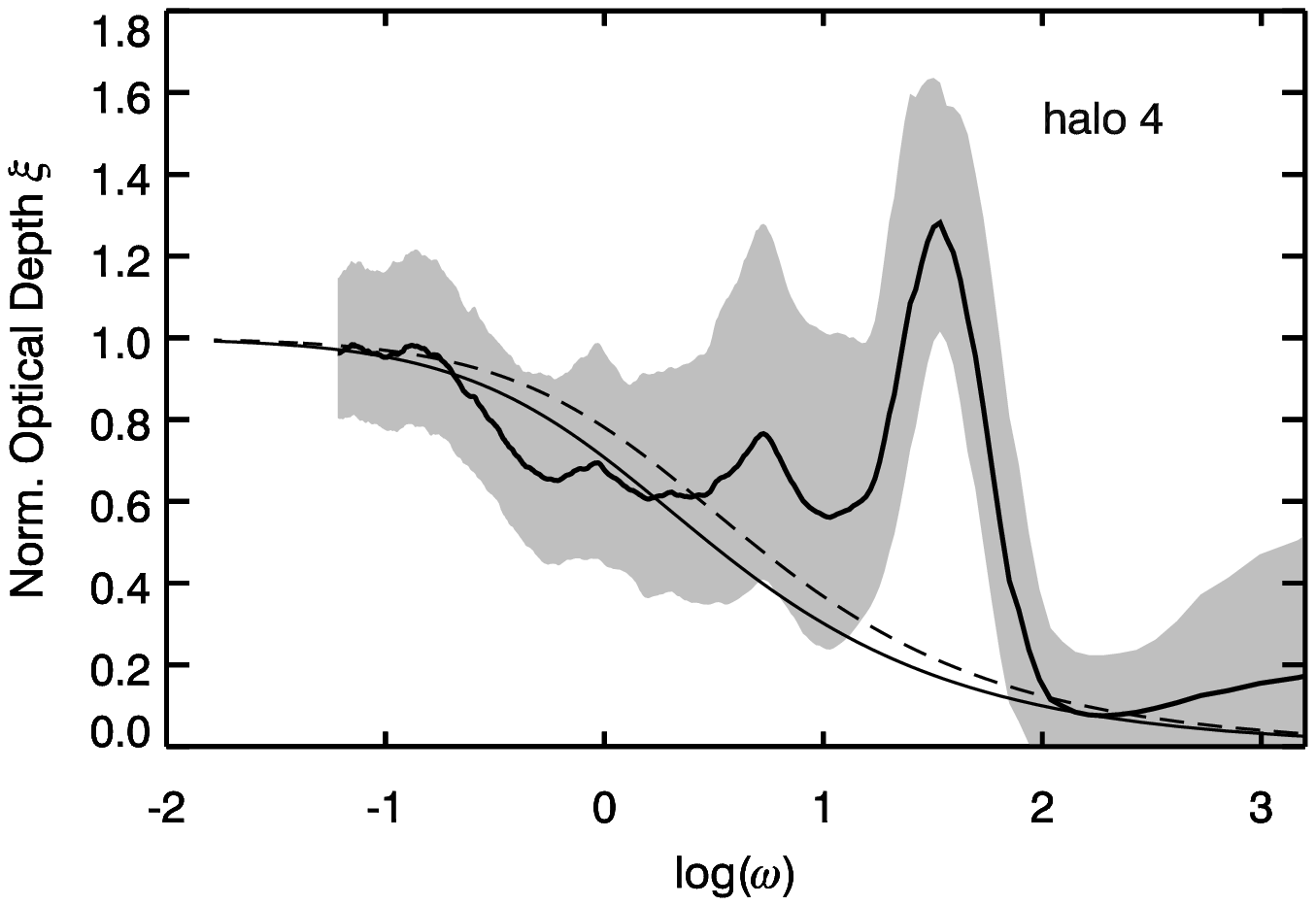}\\
\
\caption{\label{fig:exampMovAvg} Mean normalised optical depth profiles obtained using a moving
average smoothing with a kernel size of $\Delta \log(\omega) = 0.5$ for two halos taken from the
20 most massive halo sample at redshift
$z=4$ (black solid line with grey shading). The mean halo profiles are determined using 100 lines of sight
and the shaded area gives the $1\sigma$ standard deviation. The analytical proximity effect model is
given as the smooth black line, while the fitted profile is indicated by the dashed line. The left panel
shows a halo without strong intervening density features and closely following the analytic model.
The right panel illustrates a halo having strong density features in its vicinity, causing strong deviations
from the analytical form.}

\end{figure*}

In P1 we found the observed $\xi$ to fluctuate strongly around the analytical proximity effect
profile. Any fit of Eq. \ref{eq:P2strengthParam} is thus biased by these large
fluctuations which arise from the presence of strong absorbers along the line of sight. 

In order to obtain a smoother $\xi$-profile, the wavelength scale of the spectrum is transformed into
the $\log \omega$-scale and a boxcar smoothing (also known as moving average) with the size of 
$\Delta \log \omega$ is applied to the transmission spectrum. Analogously to the method given 
above, we then calculate the effective optical depth in each smoothed pixel and determine the 
normalised optical depth $\xi$. This results in smooth $\xi$-profiles allowing the identification of
areas dominated by strong absorption systems. 

In Fig. \ref{fig:exampMovAvg} two examples of mean $\xi$-profiles taken out of the 20 most massive halos
sample at $z=4$ are given. A $\Delta \log \omega = 0.5$ is found to yield a good balance between 
smoothing and retaining structure in the $\xi$-profile. For the mean $\xi$-profiles
shown in Fig. \ref{fig:exampMovAvg}, 100 lines of sight have been constructed randomly around 
the DM halos.
The two examples illustrate a halo without any signs of intervening strong absorption systems (left panel)
and one where a strong density feature is located near the host halo at $\log \omega = 1.5$ (right panel). 
From the
halo without strong nearby systems it becomes evident that the profile closely follows the analytical one.
A fit to this smooth $\xi$-profile is now very robust, since the functional form of the profile is well 
constrained and the fit of Eq. \ref{eq:P2strengthParam} is not solely
determined by a small number of data points. 
The same is true for cases with strong intervening absorption systems, as long
as the underlying smooth profile of Eq. \ref{eq:P2xiOmega1} is still visible. 
Strong deviations from the analytical form however, as presented in our second example,
are large enough to prevent a clear identification of the smooth analytical profile and the obtained
strength parameters have to be treated with caution.

\section{Results}
\label{sec:p2_results}

\subsection{Null hypothesis: Random locations}

In order to assess whether large scale over-densities affect the proximity effect
profile and the associated strength parameter, we first establish results 
that are unaffected by such large scale density
features. A sample of 500 randomly selected lines of sight originating at random
points in the simulation will serve as a null hypothesis. 
No noise is added to the spectra. However we will use this sample in the next section to 
discuss the effect of detector noise on our results.

\begin{figure*}
\centering
\includegraphics[width=0.48\textwidth]{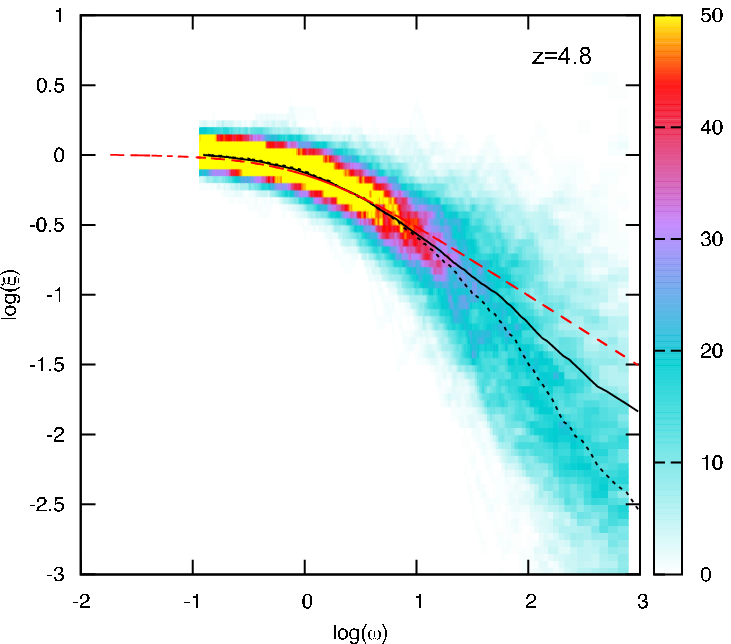}
\includegraphics[width=0.48\textwidth]{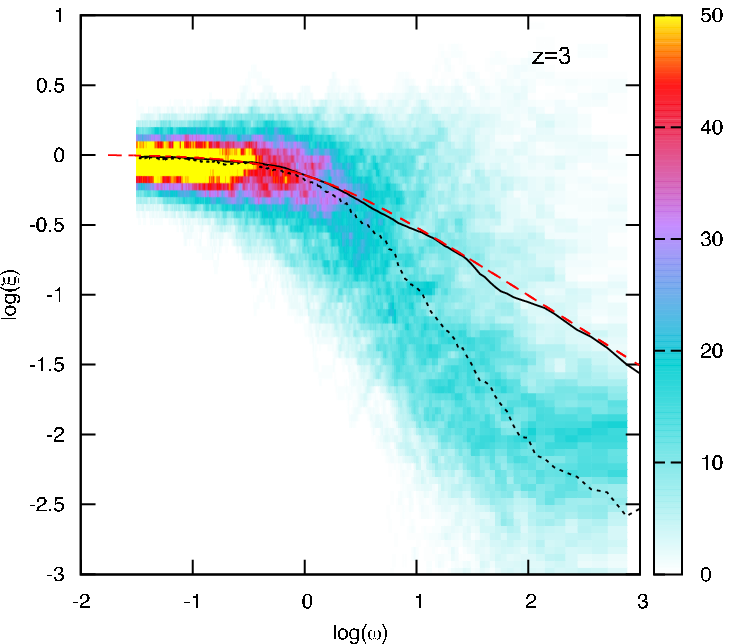}\\

\caption{\label{fig:nullHypo}Frequency of the normalised optical depth $\xi$ using $\Delta \log \omega = 0.5$ 
calculated from 500 random lines of sight drawn from our simulation box, as a function of $\omega$. 
The origins of these lines of sight were chosen not to be centred on any specific haloes, but on random 
points in the box. The left panel gives results for $z=4.8$ and the right panel $z=3$. 
The dashed line marks the
input model used for generating the spectra including the proximity effect. The black solid line marks the
mean $\xi$-profile, whereas the dotted black line marks the median profile.}

\end{figure*}

In Fig. \ref{fig:nullHypo} we discuss results obtained for redshifts $z=4.8, \textrm{ and } 3$.
For each line of sight, the normalised optical depth $\xi$ was calculated using the
method described in Sect. \ref{sec:newProxMeth}. For each redshift we determine the mean and
median $\xi$ profile, as well as the $\xi$ probability distribution as a function of $\omega$.
The $\xi$ probability distribution is shown in colour coding in Fig. \ref{fig:nullHypo}.

At $z=4.8$ the mean profile follows the analytic proximity effect model very well, with
just a slight increase in its slope towards higher $\omega$ values. Note however that
at large $\omega$ values of $\log \omega > 2$, $\xi$ can only be poorly determined 
due to a very small number of pixels contributing to the $\Delta \omega$ bins. The median
profile as well follows the input model up to $\log \omega \sim 0.5$. However
at $\log \omega > 0.5$ the median profile steepens strongly and starts to deviate from the input model 
and the mean profile. This indicates a growing asymmetry in the $\xi$ distribution with
increasing $\omega$ (approaching the QSO), resulting in the growing discrepancy between
the mean and median profiles. Considering the $\xi$ distribution function, 
the increasing skewness of the distribution
becomes apparent. For $\log \omega < 1$, the width of the distribution stays
constant and appears symmetrical and normally distributed in the logarithmic scale.
This indicates a log normal distribution of $\xi$ values. However at larger $\omega$
values the distribution starts to widen up and the peak in probability shifts towards 
lower values of $\log \xi$, moving away from the expectations of the analytical formalism. 

For redshifts $z=4$ and $z=3$ a similar picture emerges. The $\xi$ distribution
widens as a function of redshift, with an increase in its variance with decreasing
redshift. The mean profile however always regains the input model well.
However the discrepancy between the median profile and
the mean at high $\omega$ values increases with decreasing redshift. The median
profile becomes steeper and steeper, indicating a growing skewness of the
$\xi$ distribution at high $\omega$. However at low $\omega$, the $\xi$ distribution
stays symmetric and continues resembling a log normal distribution up to 
$\log \omega \sim 0$. At $z=2$, the $\xi$-distribution shows a strongly increasing
variance, dominating over the signal of the input model completely. We will therefore
not consider results from $z=2$ in this work. 

The increase in variance is dominated
by two factors. On the one hand the universe evolves and the growth of structure 
increases with decreasing redshift. This introduces stronger density contrasts between
under- and overdense regions. On the other hand, the IGM becomes more transparent
with decreasing redshift due to the increasing UV background and cosmic expansion 
reducing the mean density of the universe. Therefore at low redshifts, the Ly$\alpha$
forest traces denser structures than at high redshift.

From the simple test of this subsection we conclude that with randomly selected lines of sight of 
random origins, the input proximity
effect model can be regained with the mean $\xi$ profile. In our sample this is valid
for $z \geq 3$. The median profile however deviates more and more from the mean profile with 
decreasing redshift and cannot be used to measure the UV background from the proximity effect
with the current analytical model, which is formulated for a mean IGM.

\begin{figure*}
\includegraphics[width=0.48\textwidth]{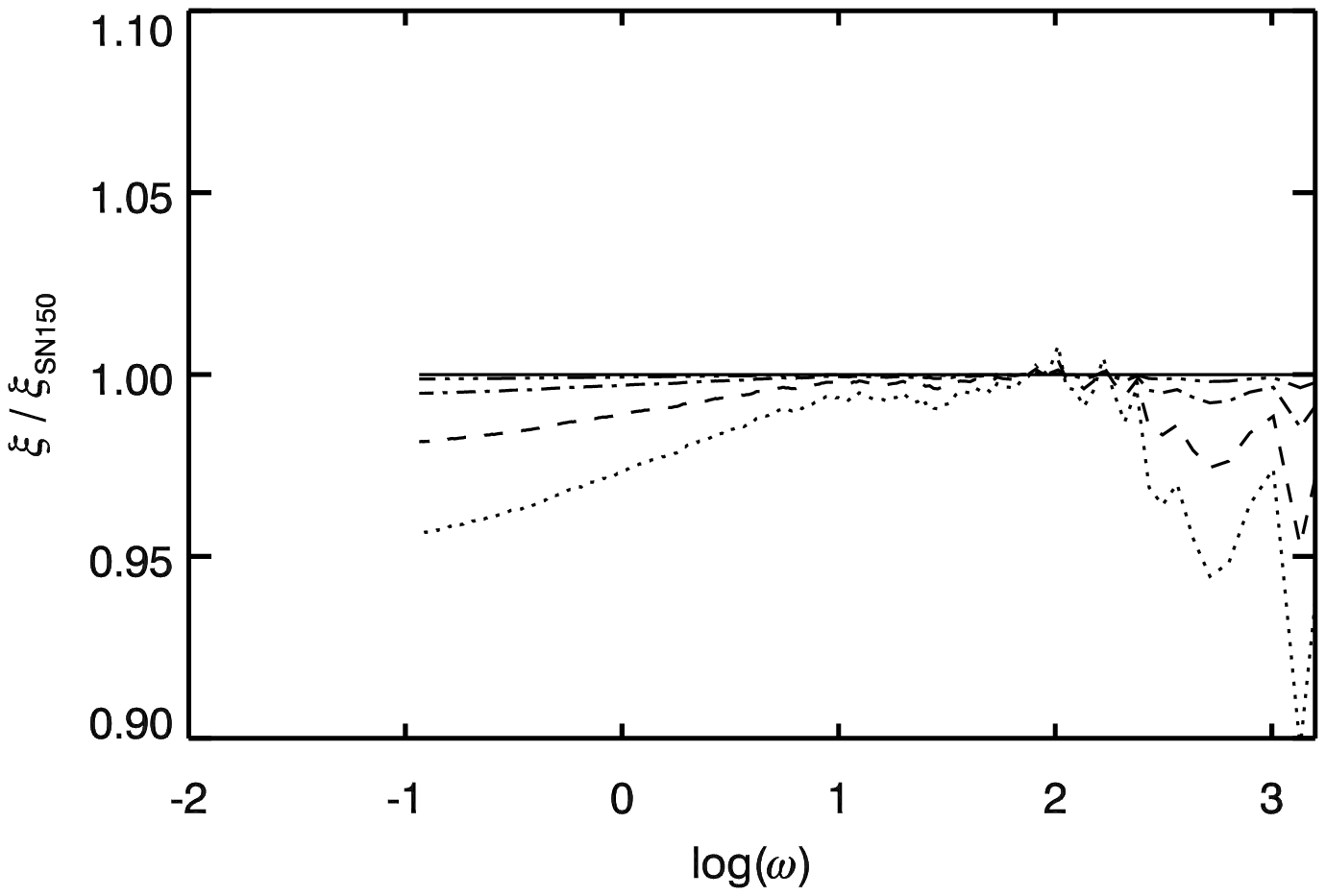}
\hfill{}%
\includegraphics[width=0.48\textwidth]{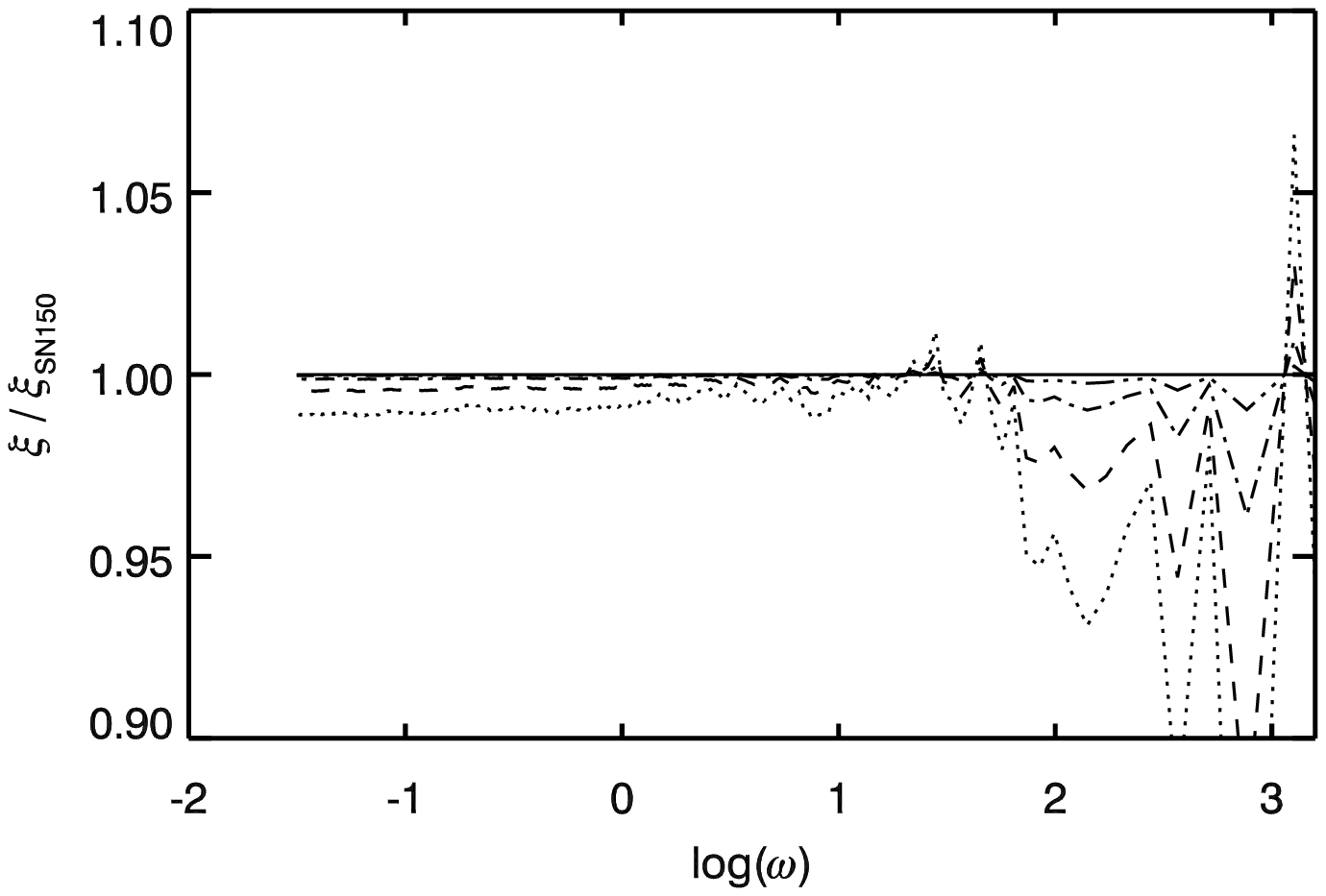}\\
\
\caption{\label{fig:nullHypoNoise} The influence of the signal to noise ratio on the mean normalised optical 
depth $\xi$ for the null hypothesis. Shown are results at redshifts $z=4.8$ (left panel) and $z=3$ (right panel).
The $\xi$-profiles have been normalised to results using a signal to noise of 150. The dotted line denotes
a S/N of 10, the dashed line a S/N of 20, the dash dotted line a S/N of 50, and the dash triple dotted line
a S/N of 100.}

\end{figure*}

\subsection{Dependence on the spectra signal to noise}
\label{sec:proxySN}
\begin{figure}
\includegraphics[width=1\columnwidth]{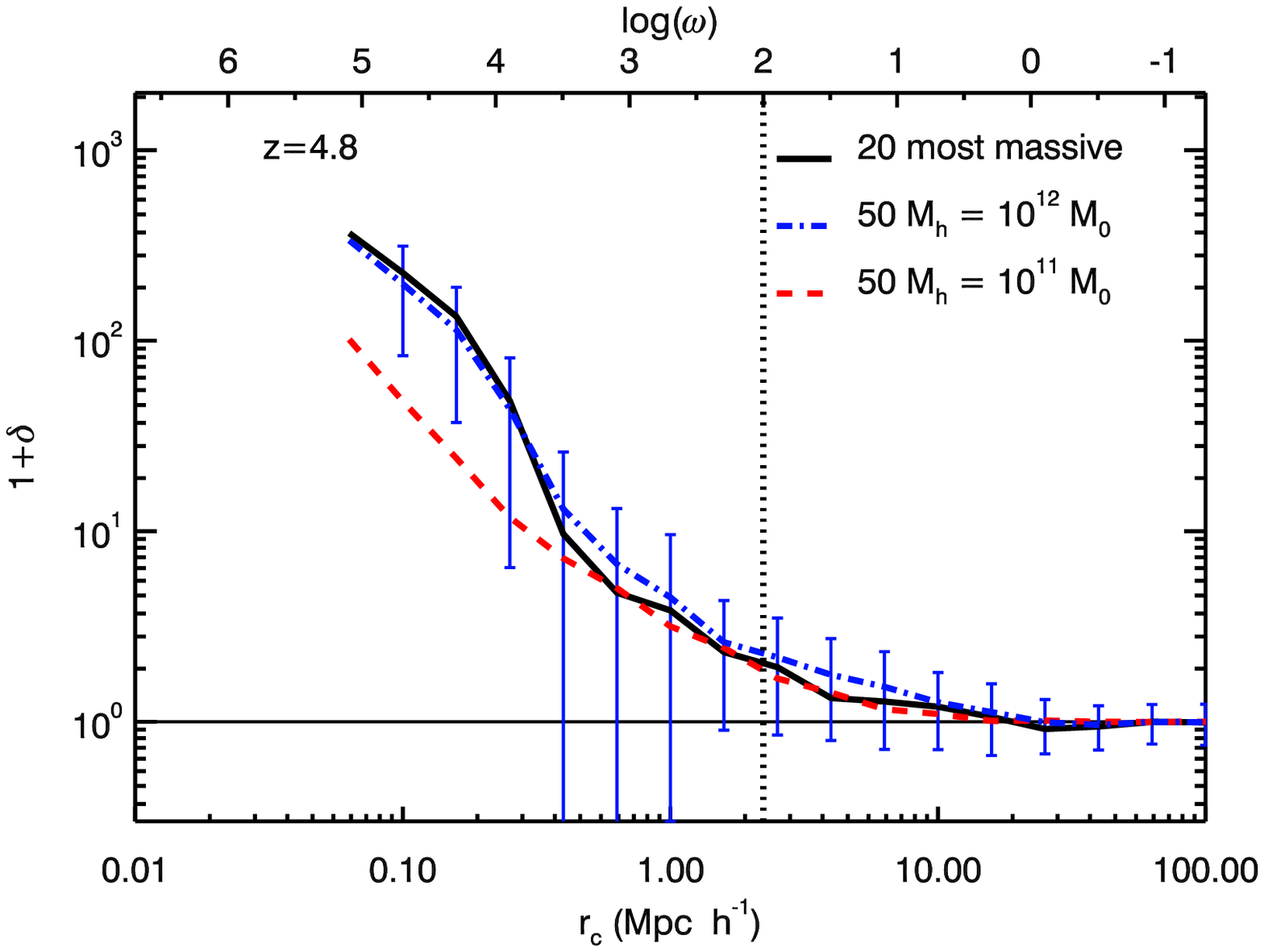}\\
\includegraphics[width=1\columnwidth]{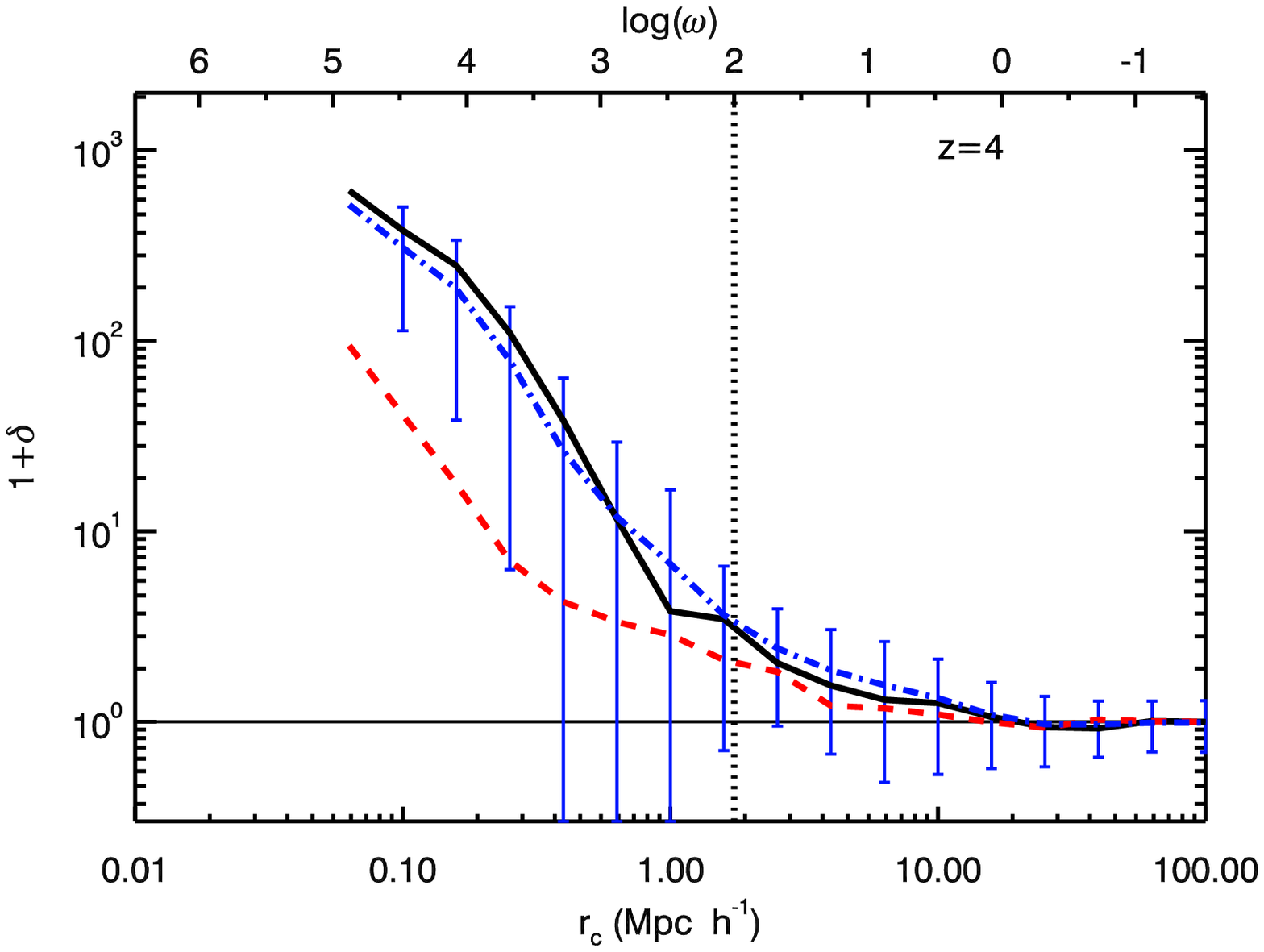}\\
\includegraphics[width=1\columnwidth]{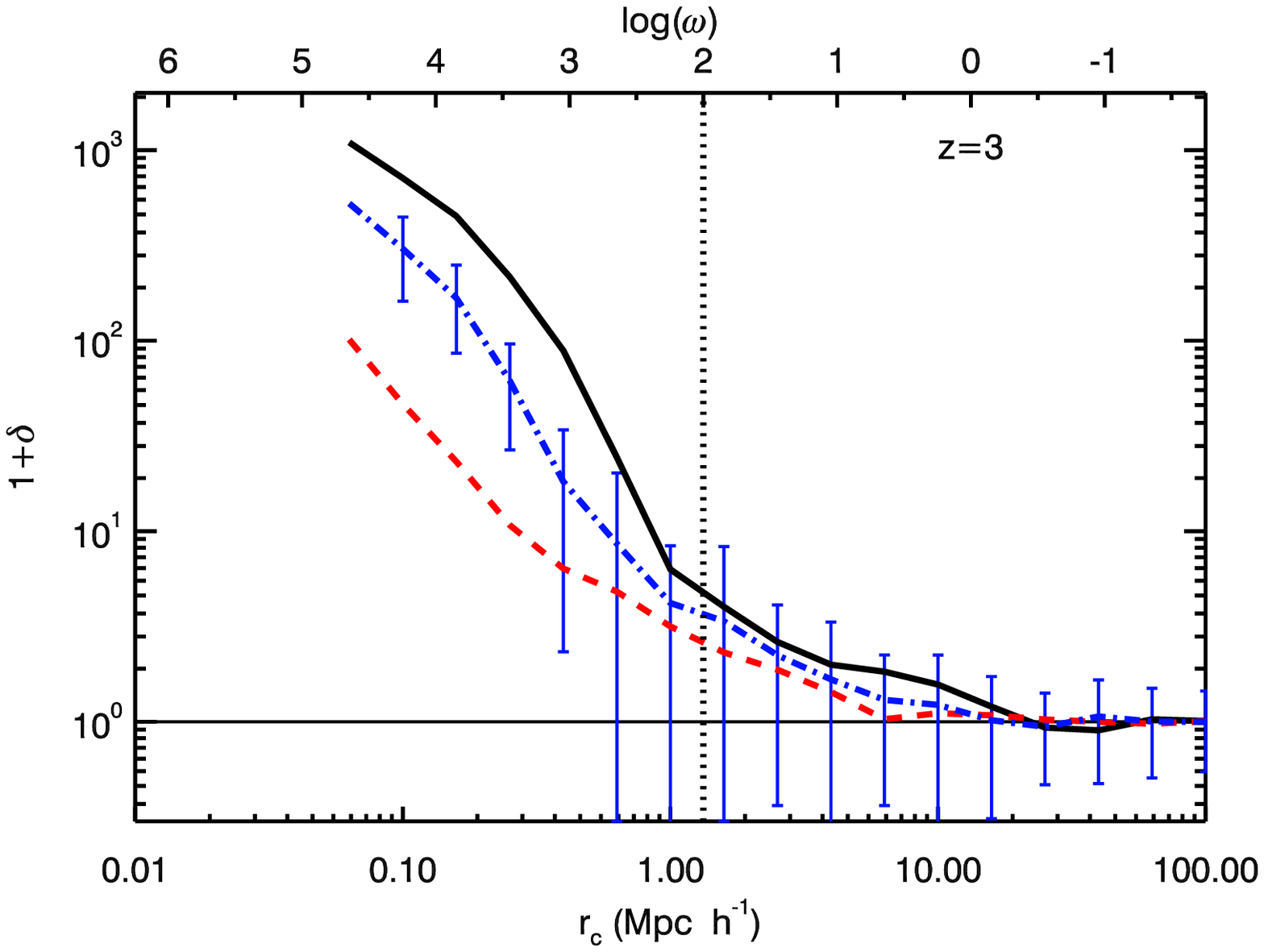}
\caption{\label{fig:meanDensity} Mean density environment around all the halos in each mass bin. The
density profiles are smoothed for better visibility. Further the corresponding $\omega$ scale is
given as reference. The black solid line denotes the mean profile
around the 20 most massive halos, the blue dash-dotted line denotes the mean around the 
$M_\mathrm{halo} = 10^{12} \; M_{\astrosun}$ halos, and the red dashed line marks the mean around the
$M_\mathrm{halo} = 10^{11} \; M_{\astrosun}$ halos. The standard deviation is given for the 
$M_\mathrm{halo} = 10^{12} \; M_{\astrosun}$ halos exemplary for the other environments. The 
first panel shows $z=4.8$, the second panel $z=4$, and the last panel $z=3$. The vertical dotted
line marks the area of $\log \omega > 2$ which was excluded in the determination of the 
strength parameter.}

\end{figure}

\begin{figure}
\includegraphics[width=1\columnwidth]{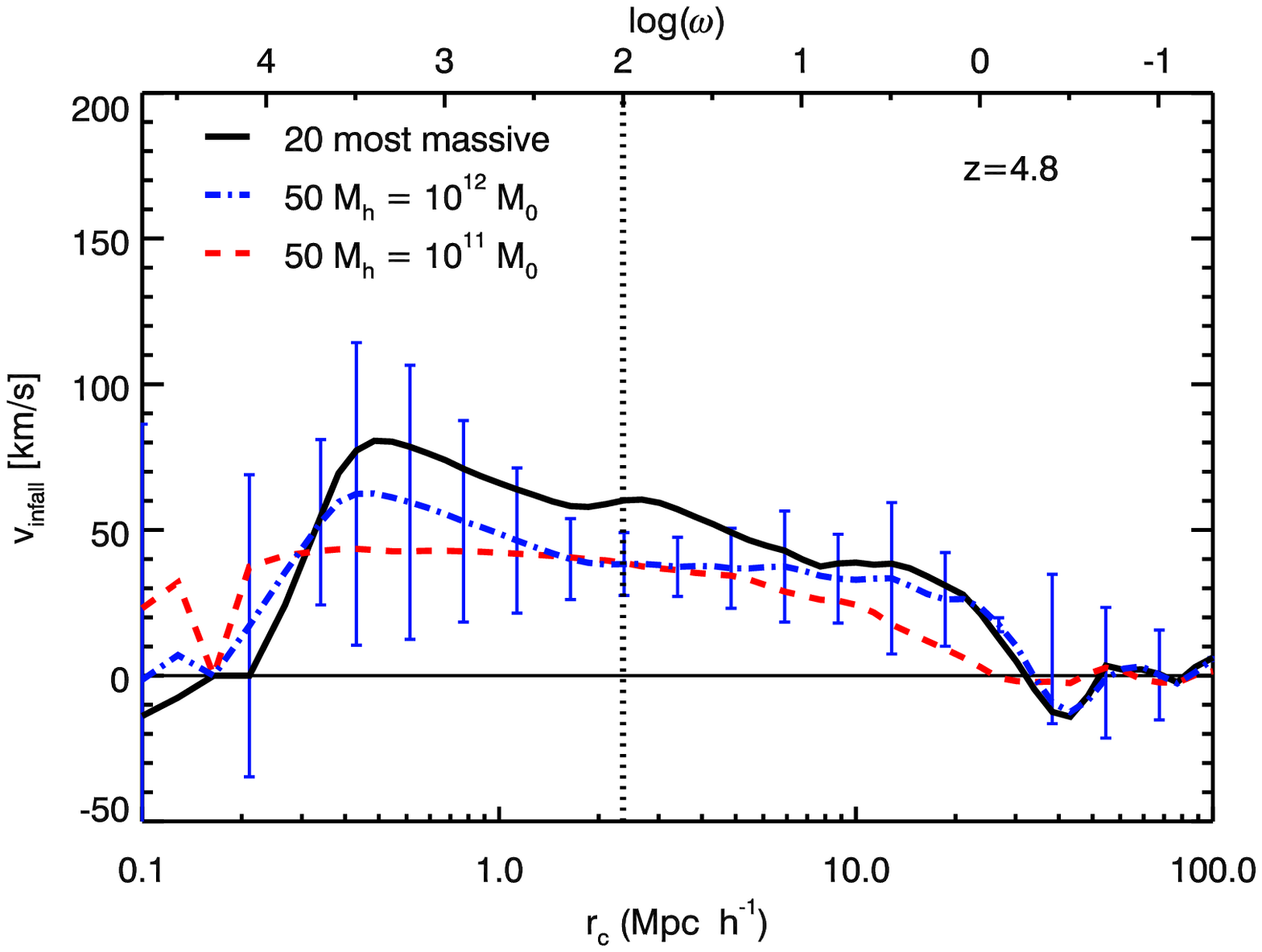}\\
\includegraphics[width=1\columnwidth]{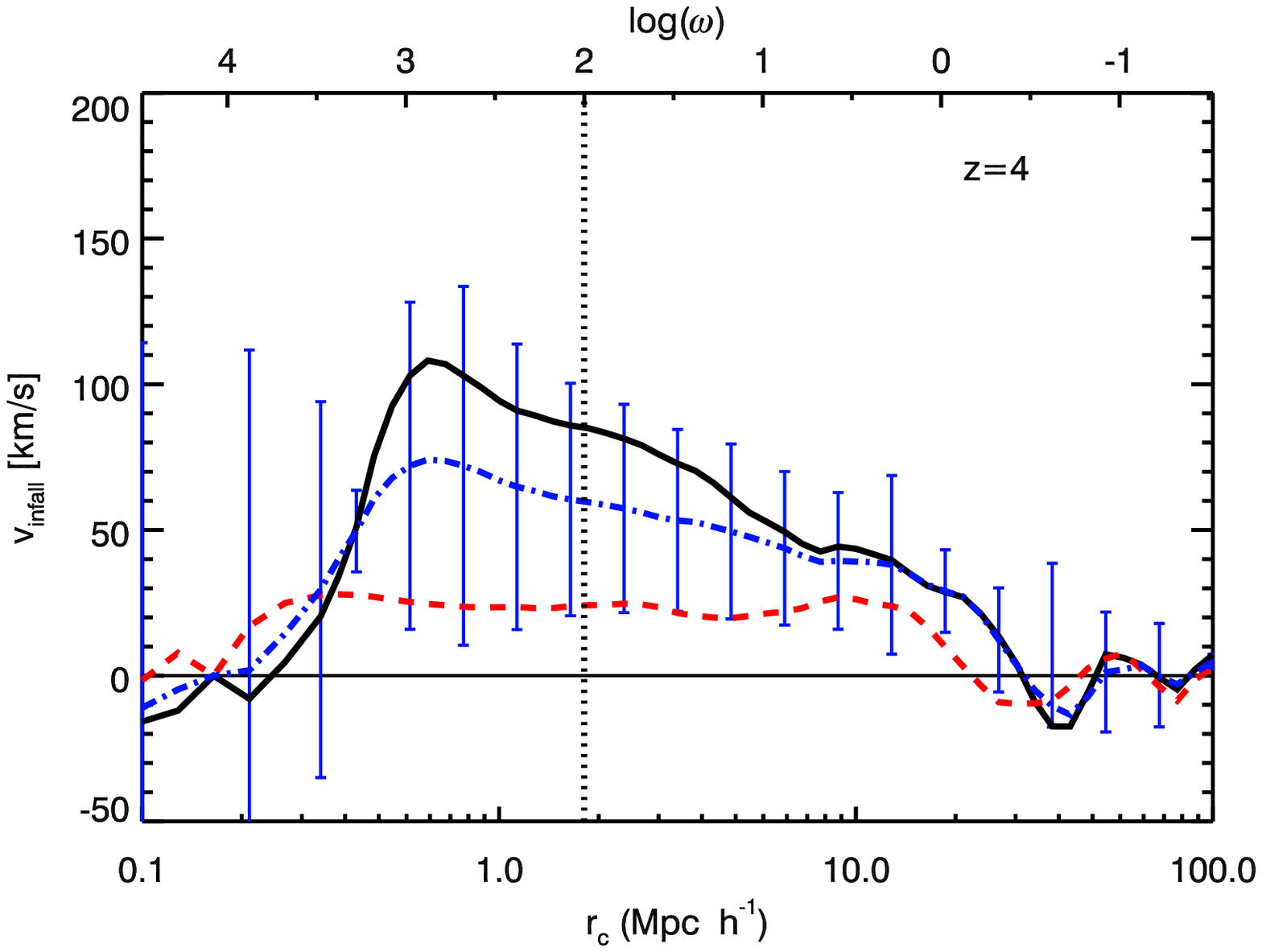}\\
\includegraphics[width=1\columnwidth]{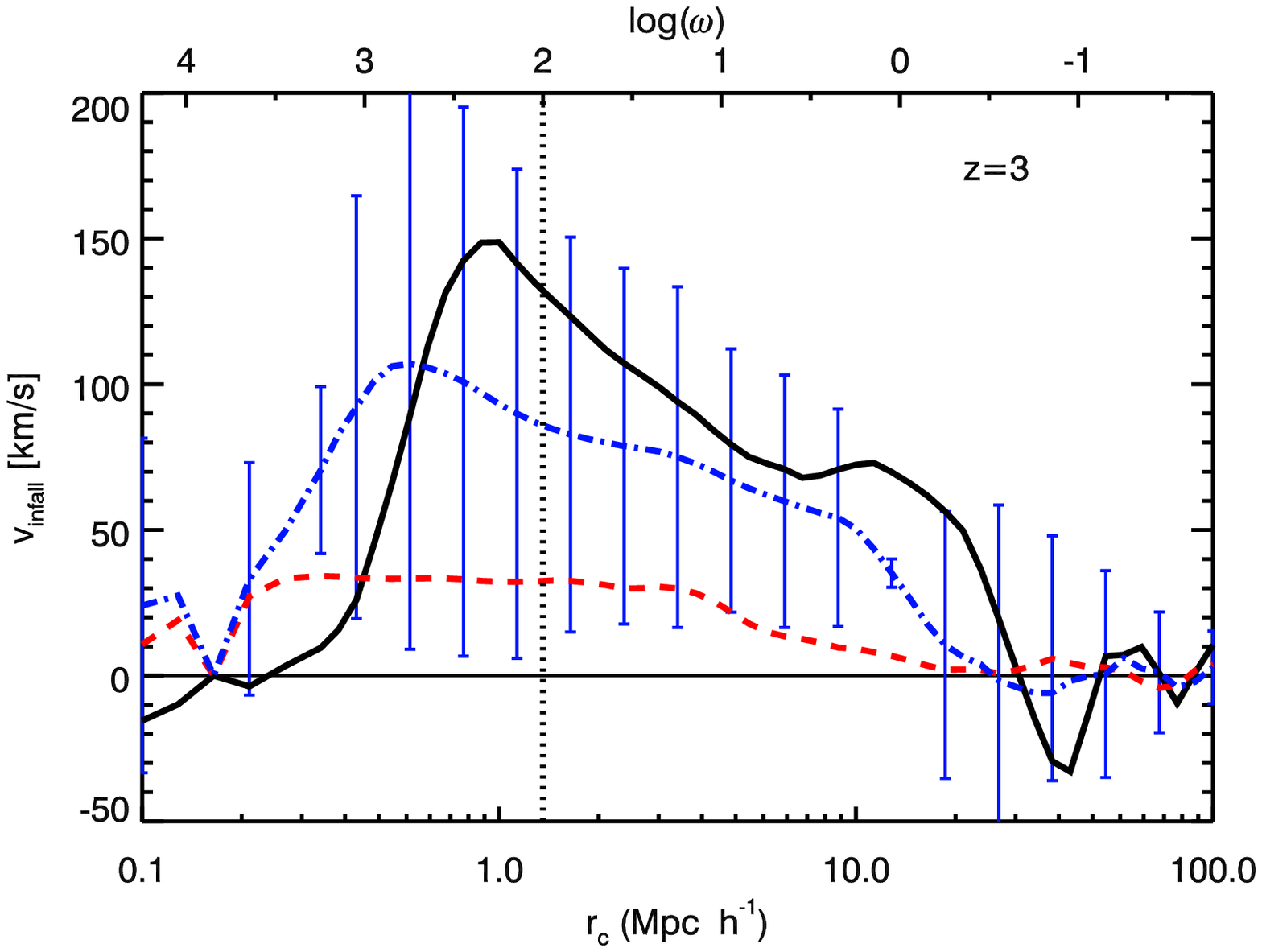}
\caption{\label{fig:meanVel} Mean infall velocity around all the halos in each mass bin. The
infall velocity profiles are smoothed for better visibility. Further the corresponding $\omega$ scale is
given as reference. The black solid line denotes the mean infall velocity
around the 20 most massive halos, the blue dash-dotted line denotes the mean around the 
$M_\mathrm{halo} = 10^{12} \; M_{\astrosun}$ halos, and the red dashed line marks the mean around the
$M_\mathrm{halo} = 10^{11} \; M_{\astrosun}$ halos. The standard deviation is given for the 
$M_\mathrm{halo} = 10^{12} \; M_{\astrosun}$ halos exemplary for the other environments. The 
first panel shows $z=4.8$, the second panel $z=4$, and the last panel $z=3$. The vertical dotted
line marks the area of $\log \omega > 2$ which was excluded in the determination of the 
strength parameter.}

\end{figure}

We now want to discuss the effect of detector noise on the mean proximity effect
profiles. Using the 500 lines of sight from the null hypothesis sample, noise is added
to each spectrum using a signal to noise (S/N) of 10, 20, 50, 100, and 150. For each
S/N sample we determine the mean $\xi$ profile and normalise it to the S/N$=150$ results.
The influence of noise on the proximity effect signature is shown in Fig. \ref{fig:nullHypoNoise},
where the normalised $\xi$ profiles are shown for $z=4.8$ and
$z=3$ as a function of the signal to noise.

The largest influence noise has on the profile is at high $\omega$ values where only
a small number of pixels contribute to $\xi$. However for $\omega$ values below $\log \omega
\sim 1.5$, the strong influence of the noise on the profile disappears. At $z=4.8$ an increase
of noise effects can be noted at $\log \omega < 0$. There the results
with a S/N of 10 produce deviations of up to 5\% from high signal to noise spectra. However already
with a signal to noise of 50, the $\xi$ profile is regained with an accuracy of less than 1\%. 
At $z=3$ convergence with the low noise results for $\log \omega <  1.5$ is already achieved with
a signal to noise of 20. In order to resolve the profile within 1\% accuracy at higher $\omega$ values, 
a signal to noise of 50 is needed. 

One has to keep in mind though that these results only apply to the combined sample and not
to individual lines of sight. However this shows that with our approach of not considering noise in the
spectra, high signal to noise results are very well approximated.

\subsection{Large scale environment}
\label{sec:p2LargeScaleEnv}

We now want to characterise the large scale density and velocityÊenvironment around our QSO host halos.
It has been shown by \citet{Prada:2006jk} and \citet{Faucher-Giguere:2008gf} that the mean
density profile around massive halos does not reach the mean cosmic density before a halocentric
distance of comoving
$r_c \sim 10 \;  h^{-1} \textrm{ Mpc}$. For smaller radii the mean DM density around halos 
can be up to 10 times above the mean density for halocentric distances of comoving $r_c
\sim 1 \;  h^{-1} \textrm{ Mpc}$. For even smaller radii the density profile is governed by the
density profile of the host halo. A similar behaviour has been observed by P1. 
\citet{Faucher-Giguere:2008gf} have further determined the bias caused by Doppler shifting of 
absorption lines due to infalling gas in proximity effect measurements. The additional bias is 
found to be similarly important to the effect of overdensities. This effect is naturally included in our
synthetic spectra taken from the simulations.

These large scale features are caused by large scale modes perturbing the density
distribution at scales of several Mpc. These large modes still experience linear growth and increase
with decreasing redshifts. Since halos with very large masses form preferably where these
large modes show a positive amplitude, an increase in density above the mean cosmic
density is expected around very massive halos. Since low mass halos also form in regions where
there is less power on large scales, the density profiles are expected to be less influenced.

In Fig. \ref{fig:meanDensity} the smoothed and radially averaged overdensity 
profiles of our halo sample are shown as a function of distance and halo mass. A consistent picture to previous 
findings emerges. At $z=4.8$ the
profiles show a steep decline in density up to comoving distances of $r_c \sim 0.4 \;  h^{-1} \textrm{ Mpc}$
for the massive halos. This region directly shows the density profile of the host halo. For larger distances the 
slope with which the density decreases becomes shallower and a transition between the host halo profile 
and its surrounding density environment is seen. The profile then reaches the mean 
cosmic density at a distance above $r_c > 10 \;  h^{-1} \textrm{ Mpc}$, independent of halo mass. 

For $z=4$ the picture is similar, however differences between the low mass halos and the more
massive ones increase. The most massive halos themselves now leave their imprint up to a
radius of $r_c \sim 1 \;  h^{-1} \textrm{ Mpc}$, whereas the direct influence of the low mass host halos 
ends at $r_c \sim 0.3  \;  h^{-1} \textrm{ Mpc}$. The low mass halos show a less pronounced large scale overdensity
with a shallower decline in density than the more massive halos. The transition
point where the density profile reaches the mean cosmic density is again found at a radius of around
$r_c \sim 10  \;  h^{-1} \textrm{ Mpc}$. Furthermore the large scale density around the halos shows slightly larger
values than at $z=4.8$, which is due to the ongoing structure formation and the linear growth of the 
corresponding large wavelength modes.

At $z=3$ the differences between the various halo masses further increases. The transition point
where the profile reaches the mean cosmic density starts to show a dependency on halo mass. The
20 most massive halos reach the mean density at a radius of $r_c > 20  \;  h^{-1} \textrm{ Mpc}$, whereas the
low mass halos already reach it at $r_c > 7  \;  h^{-1} \textrm{ Mpc}$. Furthermore the large scale overdensity 
has grown in density. We therefore expect any
influence of this large scale overdensity on the proximity effect measurement to increase with
decreasing redshift. Additionally we do not expect a dependency of a possible bias
with halo mass at high redshifts, since no large differences between the density profiles are seen. 
However at low redshifts, differences due to stronger large scale
overdensities with increasing halo mass may leave an imprint in the proximity effect profile. 

For completeness we show the mean infalling velocities as a function of distance to the host
for each halo mass bin in Fig. \ref{fig:meanVel}. Analogously to the density environment, the infall 
velocities grow with decreasing redshift due to the onset of structure formation. The infall velocity
profiles for the 20 most massive halos and the $M_\mathrm{halo} = 10^{12} \; M_{\astrosun}$ sample
behave similarly at all redshifts. Moving away from the halo, the infall velocity rises until it reaches its
maximum at the position where the density profile shows a transition between the host halo and its
surrounding density environment. At larger radii the infall velocity steadily decreases until no 
systematic infall towards the halo is seen anymore. The radii where the systematic infall vanishes
is about the size of the large scale overdensity itself. In the case of the low mass halo though, the
infall velocity remains constant at low velocities over the whole large scale overdensity.

\subsection{Proximity effect strength as function of halo mass}

\begin{figure}
\includegraphics[width=0.48\textwidth]{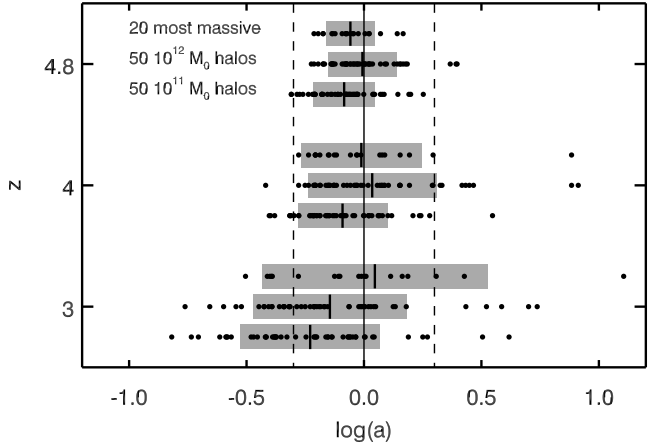}
\
\caption{\label{fig:aParamMassMeanMedian} The $a$ parameter distribution as a function of halo mass and
redshift. Each black dot represents the $a$ parameter obtained using the mean halo profile
of 100 lines of sight. 
For each redshift the first line shows results from the 20 most massive halos
sample, the second line represents the $M_\mathrm{halo} = 10^{12} \; M_{\astrosun}$ sample, and the third
line the $M_\mathrm{halo} = 10^{11} \; M_{\astrosun}$ sample. The bold black lines indicate the sample's mean
$a$ parameter, while the grey shaded areas indicate the corresponding
$1\sigma$ standard deviation. The solid line marks the
fiducial model, while the dashed lines indicate deviations of a factor of two in the UVB measurements.}

\end{figure}

We now want to test whether the proximity effect strength parameter $a$ shows any
dependency with the host halo mass. For each halo the mean $\xi$ profile is
calculated from 100 lines of sight centred on the halo position. 
Then Eq. \ref{eq:P2strengthParam} 
is fitted to the mean profile and a proximity effect strength parameter $a$ is obtained for each halo
in our sample. Any strength parameter with $\log a > 0$ results in an overestimation of the UV background
photoionisation rate in real measurements, whereas a $\log a < 0$ resembles an underestimation.

In Fig. \ref{fig:aParamMassMeanMedian} each halo's mean strength parameter $a$ is marked as a function of
redshift and halo mass. The $a$ parameter distribution is thus obtained for each halo mass bin, and its
mean and standard deviation are marked by the vertical bold line and grey area in the plot. 
At each redshift, the mean
values of the distribution show no clear dependency with the host halo mass within the $1\sigma$ fluctuations. 
For redshift $z=4.8$, the different
halo bins yield similar mean values and standard deviations. Considering the variance
in the distribution, the input model is regained for all the halo mass bins at $z=4.8$.

This picture does not change with redshift. Against the expectations motivated by the overdensity profiles
discussed in Sect. \ref{sec:p2LargeScaleEnv}, the data shows no dependency on the halo mass. Even
at $z=3$, where the mean large scale overdensity was found to be lower around lower mass halos than around
high mass ones, no significant dependency is seen. The mean values are thus consistent with the input 
model within $1\sigma$. The halo to halo variance increases with decreasing redshift and the strength
parameter distribution broadens. At $z=4.8$ the standard deviation of the whole
halo sample is $\sigma(\log a)=0.08$, while at $z=4$ we find $\sigma(\log a) = 0.20$ and 
$\sigma(\log a) = 0.36$ for $z=3$. A similar increase in the variance has been previously observed by 
\citet{DallAglio:2008ua} for individual lines of sight of a high-resolution QSO sample. 

From these results we conclude that the proximity effect measurements do not show any host
halo mass dependent bias. The variance in the signal between different halos increases with decreasing
redshift and tighter constraints are obtained with higher redshift objects. However even at low
redshifts, the mean strength parameter regained the input value within a factor of less than two. 

\subsection{Influence of the large scale overdensity}

\begin{figure}
\includegraphics[width=0.9\columnwidth]{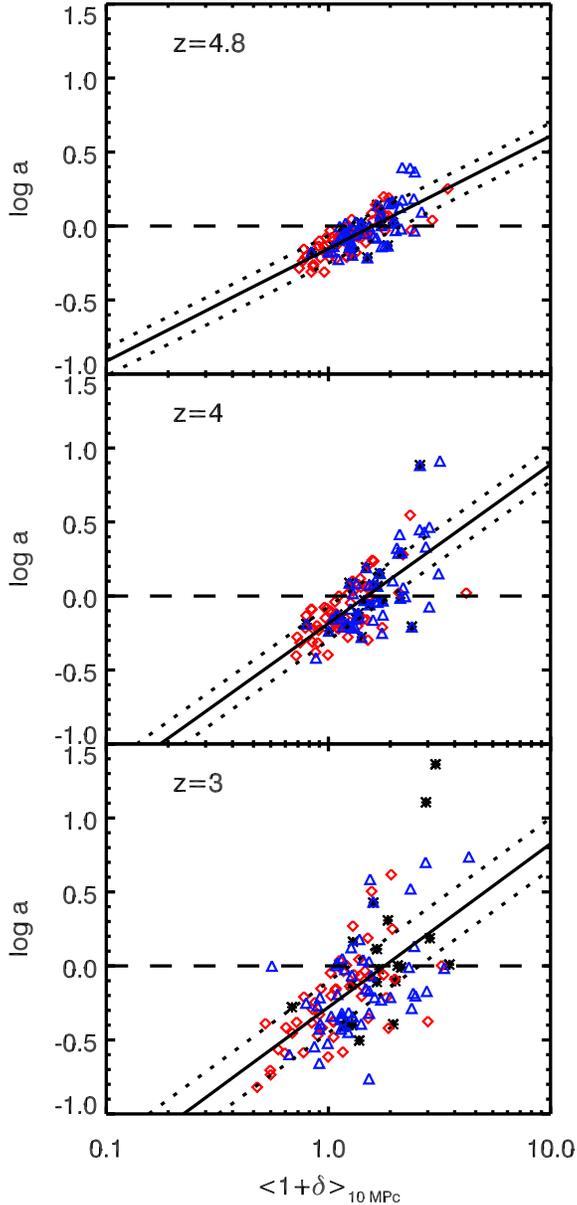}
\
\caption{\label{fig:intDensAParam} Proximity effect strength parameter as a function of the mean
overdensity in a shell of $1$ to $10 \;  h^{-1}  \textrm{ Mpc}^{-1}$ comoving radius 
around the 20 most massive halos (black asterisks),
$M_\mathrm{halo} = 10^{12} \; M_{\astrosun}$ halos (blue triangles), and 
$M_\mathrm{halo} = 10^{11} \; M_{\astrosun}$ halos (red diamonds). 
The black dashed line marks $a=1$. The solid line represent linear
regressions obtained from the dataset where the dotted lines mark the $1\sigma$ upper
and lower boundaries.}
\end{figure}

In the previous section we have shown that the proximity effect strength parameter
is not affected by the host halo mass. However a large scatter around the mean strength
parameter is seen in each mass bin, which increases with decreasing redshift. We now
want to establish the physical cause of the scatter, and therefore check if this scatter 
is connected with the density enhancement seen on large scales of up to comoving radii of
$r_c = 10 \;  h^{-1}  \textrm{ÊMpc}$. 

For each halo, the mean overdensity profile is calculated using the density distribution 
along each line of sight. Then the mean over-density $\left<\delta \right>_R$ between 
the radii $r_\mathrm{1}$ and $r_\mathrm{2}$ is calculated from the mean density
distribution around the halo using
\begin{equation}
\left<\delta \right>_R = \frac{1}{r_\mathrm{2} - r_\mathrm{1}} \int_{r_\mathrm{1}}^{r_\mathrm{2}} \delta(r) \; \mathrm{d} r.
\end{equation}
We choose $r_\mathrm{1} = 1 \;  h^{-1}  \textrm{ Mpc}$ comoving in order to exclude the influence of the
host halo on the density profile, since we are only interested in the large scale density distribution. 
The integral extends to comoving $r_\mathrm{2} = 
10 \;  h^{-1}  \textrm{ Mpc}$ which is the characteristic size of the large scale density enhancements 
(see Sect. \ref{sec:p2LargeScaleEnv}).

In Fig. \ref{fig:intDensAParam} the strength parameter $a$ is correlated with the mean overdensity 
in the comoving radial interval $r_c = [1,\;10]  \;  h^{-1} \textrm{ Mpc}$. The various symbols indicate
the different halo mass samples. At all redshifts, a correlation between
the mean density in the vicinity of the QSO and the strength parameter of the mean
$\xi$-profile is seen. A linear regression 
$\log a = \log \left< 1 + \delta \right>_0 + \alpha \log \left< 1 + \delta \right>_\mathrm{10 Mpc}$ 
is derived from the data, where $\log \left< 1 + \delta \right>_0$ is the normalisation point and 
$\alpha$ is the slope of the regression. The resulting fit parameters are listed in Table 
\ref{tbl:linRegDensStrengthParam} together with the Pearson product-moment correlation coefficient 
$r$ of the data set.

\begin{table}
\centering 
\caption{\label{tbl:linRegDensStrengthParam}
Table of linear regression results of the overdensity-strength parameter relation, together
with the data's Pearson product-moment correlation coefficients $r$. The uncertainties in the fitting
parameters have been determined with the bootstrap technique.}
\begin{tabular}{@{}c@{~~~}c@{~~~}c@{~~~}c}
\hline
$z$ & $\log \left< 1 + \delta \right>_0$ & $\alpha$ & $r_\mathrm{Pearson}$ \tabularnewline
\hline 
$4.8$ & $-0.15\pm0.07$ & $0.76\pm0.08$ & $0.74$ \tabularnewline
$4.0$ & $-0.19\pm0.13$ & $1.07\pm0.16$ & $0.68$ \tabularnewline
$3.0$ & $-0.28\pm0.19$ & $1.11\pm0.17$ & $0.60$ \tabularnewline
\hline
\end{tabular}
\end{table}

Looking at the $z=4.8$ results, the data points align clearly on a power law relation. 
With a Pearson correlation coefficient of $r_\mathrm{Pearson}=0.74$ the data 
exhibits a tight correlation between the two quantities. 
The higher the mean density around a halo becomes, the stronger the proximity effect is biased towards
larger strength parameters. At $z=4.8$ a strength parameter of $a=1$ (resembling an unbiased measurement
of the photoionisation rate) is obtained for a mean overdensity of 
$\left< 1 + \delta \right>_{10 \mathrm{ Mpc}} \sim 1.6 \pm 0.35$. 
It is interesting to note that there is no strong segregation between the various halo mass bins 
along the density axis. However, the lower mass halos lie to slightly lower overdensities than
the more massive ones. All the different mass bins cover a similar range of large scale densities. 
Again this indicates that there is no distinct connection between the halo mass and the mean overdensity 
on these large scales. Hence, only the amount of matter in the greater vicinity of a halo is responsible 
for the large scatter in the strength parameter $a$ seen in the previous section. 
However, a small scatter in $a$ of $\approx 20\%$ around the linear regression remains at $z=4.8$.

Similar findings apply to the lower redshift results. The $a$ parameters are still related through a 
power law with the large scale overdensity, however the strength of the correlation drops to 
$r_\mathrm{Pearson}=0.60$ at
$z=3$. Comparing the $z=4.8$ results with the lower redshift ones shows a slight steepening of 
the relation from $\alpha = 0.76\pm0.08$ at $z=4.8$ to $\alpha = 1.11\pm0.17$ at $z=3$. 
The scatter around the power law increases to $\approx 35\%$ for $z=4$ and 
55\% for $z=3$. Further the scatter seems to spread with increasing density, however a larger halo sample 
from a larger simulation would be
needed to conclusively determine this increase in the variance. Again no bias in the strength parameter is
found for an overdensity of $\left< 1 + \delta \right>_{10 \mathrm{ Mpc}} \sim 1.5 \pm 0.52$ at $z=4$
and $\left< 1 + \delta \right>_{10 \mathrm{ Mpc}} \sim 1.8 \pm 0.77$ at $z=3$.

These results show that the large scale distribution of matter around a host halo affects the proximity 
effect strength parameter and biases the measurements of the UV background photoionisation rate.
There is a power law correlation between the mean density surrounding a QSO host and
the $a$ parameter. The bias on the UV background photoionisation rate can be determined, if 
the mean overdensity around a QSO is inferred. However, density measurements from the Ly$\alpha$
forest are degenerated with the UV background photoionisation rate. The properties of the 
UV background have to be known in order to convert optical depths into densities. 

Observationally, \citet{DOdorico:2008az} have shown 
that QSOs at $z=2.6$ are situated in regions showing a density excess on scales of 4 proper Mpc. 
Due to the dependency of the results on the UV background photoionisation rate, their results only 
constrain the mean density of the large scale overdensity to a factor of 
about 3. Nevertheless we estimate a mean density in a radius of 4 proper Mpc from their results. 
We obtain an upper value on the mean overdensity of $\left< 1+\delta \right> \sim 3.5$ and a lower
value of $\left< 1+\delta \right> \sim 1.25$. According to our results at $z=3$, this would translate into a 
bias of the strength parameter of $a = 2.1$ for an overdensity of $3.5$ and $a=0.7$ for the lower value 
on the mean density around a QSO. Remember though that our results only apply to a QSO with
a Lyman limit luminosity of $L_{\nu_\mathrm{LL}} = 10^{31} \textrm{ erg Hz}^{-1} \textrm{ s}^{-1}$.
According to P1 the bias becomes smaller for higher luminosity QSOs. 

\section{Conclusion}
\label{sec:p2_conclusions}

We have studied the effect of the host halo mass and the large scale density distribution
on measurements of the proximity effect strength parameter. The strength parameter is 
observationally used to infer
the UV background photoionisation rate. From a $64 \; h^{-1} \textrm{ Mpc}$ sized
dark-matter simulation, we picked the 20 most massive halos, 50 halos with a mass around
$M_\mathrm{halo} = 10^{12} \; M_{\astrosun}$, and another 50 halos with a mass
around $M_\mathrm{halo} = 10^{11} \; M_{\astrosun}$ at redshifts $z=4.8, \, 4, \textrm{ and } 3$.
Each halo is individually assumed to host a QSO with a Lyman limit luminosity 
of $L_{\nu_\mathrm{LL}} = 10^{31} \textrm{ erg Hz}^{-1} \textrm{ s}^{-1}$. 

Around each halo, 100 random lines of sight of the density and velocity distribution
are obtained. Using models of the IGM which we calibrated to
observational constraints, the neutral hydrogen fractions and gas temperatures along each line of sight
are inferred. To include the increase in ionisation of an additional QSO radiation field, the neutral hydrogen 
fractions are decreased proportional to geometric dilution of the QSO flux field \citep{Partl:2010qm}.
Then, for each line of sight, a Ly$\alpha$ forest spectrum is computed. From the spectrum we derive
the mean proximity effect signature with observational methods and measure the proximity effect
strength parameter $a$. Since the UV background photoionisation rate used in generating the
Ly$\alpha$ spectra is known, the strength parameter quantifies the over- or underestimation of
the UV background.

In order to assess whether the QSO host environment affects the proximity effect signal, a null
hypothesis test was performed. From the simulation box, 500 lines of sight originating at random
positions with random directions have been obtained. On each line of sight, the proximity effect 
was modelled. From these lines of sight, we obtained the distribution of normalised optical depths
$\xi$ as a function of normalised distance to the QSO $\omega$. 
From the $\xi$-distribution, the mean and median proximity effect signals were determined. The mean
profiles match the analytic model very well and the model UV photoionisation rates are regained. At large
halocentric distances, the median and the mean profiles match. However the median increasingly 
deviates from the analytic model and the mean profile with decreasing distance to the QSO. 
This indicates that the $\xi$-distribution closely resembles a log normal distribution at large distances
from the QSO, however it is increasingly skewed when nearing the source. Not only is the variance of the 
$\xi$-distribution found to increase with decreasing redshift, but its skewness increases as well. 

We further determined the mean density distribution and infall velocity
around all the halos in each mass bin as a function
of redshift. For all redshifts the density profile does not immediately reach the mean cosmic density
at the border of the dark matter halo, however it stays above the mean density up to comoving halocentric
radii of $r_c \gtrsim 10 \; h^{-1} \textrm{ Mpc}$. For redshifts $z=4.8$ and $z=4$ the size of this
large scale overdensity is independent of the halo mass. However higher mass halos show in the
mean a larger overdensity than lower mass halos. At redshift $z=3$ the size of the large
scale overdensity is smaller for the lower mass sample than for the more massive ones. 
The velocity field shows a mean infall up to distances of $30 \; h^{-1} \textrm{ Mpc}$ around the
halos, reaching for the most massive halos $80$, $100$, and $150 \textrm{ km/s}$ for redshifts
$z=4.8, \; 4,$ and $3$, respectively. Overdensity and infall velocities act together in the derived
bias of the proximity effect.

The mean strength parameter per halo mass bin does not show a dependency with the halo mass.
For each halo mass bin, a mean strength parameter which is consistent with the input model is regained
within $1\sigma$ standard deviation. However the halo to halo variance is found to increase with
decreasing redshift from $\sigma(\log a) = 0.08$ at $z=4.8$ to $\sigma(\log a) = 0.36$ at $z=3$.

The strength parameter is found to correlate with the mean density measured in a shell
of comoving $1 - 10 \; h^{-1} \textrm{ÊMpc}$. We fit a power law to this correlation.
Regions with a mean density below the cosmic mean show a stronger
proximity effect than regions with densities above the cosmic mean. The correlation is tightest at $z=4.8$
and the scatter around the power law increases with decreasing redshift. Furthermore, the power law is found to
steepen slightly with decreasing redshift. From these results we find that an unbiased UV background 
photoionisation rate can only be obtained if the mean density around the QSO is between  
$\left< 1 + \delta \right>_{10 \mathrm{ Mpc}} \sim 1.5 \pm 0.52$ and  $1.8 \pm 0.77$. 

If a possibility exists to determine the mean density around a QSO, the bias arising from the large scale
overdensity can be corrected for. However density measurements from Ly$\alpha$ forest spectra are
degenerate with the UV background photoionisation rate. Due to this degeneracy, previous measurements 
of the density distribution around QSOs, such as \citet{DOdorico:2008az}, have only been able to 
constrain the mean density up to a factor of 3.
By inferring the bias of the proximity effect at $z\sim3$ according to their density measurements,
the strength parameter ranges from $a=0.7$ to $a=2.1$
for a QSO with a Lyman limit luminosity of 
$L_{\nu_\mathrm{LL}} = 10^{31} \textrm{ erg Hz}^{-1} \textrm{ s}^{-1}$.

\section*{Acknowledgments}
We are grateful to George Becker for kindly providing us the data of the flux probability
distribution functions. Further we appreciate stimulating discussions with 
Tae-Sun Kim, Lutz Wisotzki and Aldo Dall'Aglio.
AP acknowledges support in parts by the German Ministry
for Education and Research (BMBF) under grant FKZ 05 AC7BAA. 
The simulations used in this work have been performed in the Marenostrum supercomputer 
at the BSC Barcelona, the HLRB2 ALTIX supercomputer at LRZ Munich and the Juropa 
supercomputer at NIC Juelich.
GY would like to thank the MICINN (Spain) for financial support through research grants
FPA2009-08958 and  AYA2009-13875-C03-02.

\bibliographystyle{mn2e}
\bibliography{literature}

\label{lastpage}

\end{document}